\documentclass[lettersize,journal]{IEEEtran}
\usepackage{amsmath,amsfonts}
\usepackage{algorithmic}
\usepackage{algorithm}
\usepackage{booktabs}
\usepackage{diagbox}
\usepackage{caption}
\usepackage{multirow}
\usepackage{array}
\usepackage[caption=false,font=footnotesize,labelfont=rm,textfont=rm]{subfig}
\usepackage{textcomp}
\usepackage{subfig}
\usepackage{url}
\usepackage{verbatim}
\usepackage{graphicx}
\usepackage{cite}

\usepackage{color}
\usepackage{placeins}
\usepackage{hyperref}
\hypersetup{
    colorlinks=true,
    linkcolor=blue,
    filecolor=magenta,      
    urlcolor=cyan,
    pdftitle={Overleaf Example},
    pdfpagemode=FullScreen,
    }

\begin{document}

\title{Diffusion-Prior Split Gibbs Sampling for Synthetic Aperture Radar Imaging under Incomplete Measurements}

\author{Hefei Gao, Tianyao Huang,~\IEEEmembership{Member,~IEEE}, Letian Guo, Jie He~\IEEEmembership{Member,~IEEE}, Yonina C. Eldar~\IEEEmembership{Fellow,~IEEE}

\thanks{This research was supported by the National Natural Science Foundation of China under Grant 62171259. (\textit{Corresponding author: Tianyao Huang.})}
\thanks{Hefei Gao, Tianyao Huang and Jie He are with the School of Computer and Communication Engineering, University of Science and Technology Beijing, Beijing 100083, China. (e-mail: \texttt{hefeigao@xs.ustb.edu.cn}; \texttt{huangtianyao@ustb.edu.cn}; \texttt{hejie@ustb.edu.cn}).}
\thanks{Letian Guo is with the High Power Microwave Lab, Northwest Institute of nuclear technology, Xi'an 710024, China. (e-mail: \texttt{guoletian19901230@126.com}).}
\thanks{Yonina C. Eldar is with the Faculty of Math and Computer Science, Weizmann Institute of Science, Rehovot 7610001, Israel. (e-mail: \texttt{yonina.eldar@weizmann.ac.il}).}

}



\maketitle

\begin{abstract}
Synthetic aperture radar (SAR) imaging plays a critical role in all-weather, day-and-night remote sensing, yet reconstruction is often challenged by noise, undersampling, and complex scattering scenarios. Conventional methods, including matched filtering and sparsity-based compressed sensing, are limited in capturing intricate scene structures and frequently suffer from artifacts, elevated sidelobes, and loss of fine details. Recent diffusion models have demonstrated superior capability in representing high-order priors; however, existing diffusion-based SAR methods still yield degraded reconstructions due to oversimplified likelihood approximations in guided sampling. In this work, we propose a diffusion-driven split Gibbs sampling framework for SAR reconstruction, rigorously integrating measurement fidelity with learned diffusion priors. By alternately performing likelihood- and prior-driven updates via proximal sampling, this method ensures progressive convergence toward the true posterior while fully leveraging the expressive power of diffusion priors. Extensive experiments on simulated and Sentinel-1A datasets demonstrate substantial performance improvements: over 7~dB average PSNR gain in simulations, along with significant sidelobe suppression (MPLSR +2.96~dB, MISLR +11.5~dB) with respect to the best baseline result. On real-world Sentinel-1A data, the method achieves an average PSNR gain of 1.6~dB while effectively reducing artifacts and preserving scene details, including ridges, edges, and fine textures. These results underscore the potential of the adapted framework as a robust and generalizable solution for high-fidelity SAR imaging across diverse sensing scenarios. 
\end{abstract}


\section{Introduction}
\IEEEPARstart{S}{ynthetic} Aperture Radar (SAR) is an active microwave imaging system that synthesizes a large effective aperture by coherently processing echoes from a moving small antenna, thereby achieving high-resolution imaging in both azimuth and range through aperture synthesis and pulse compression techniques\cite{kovaly1976synthetic}. Operating independently of daylight and capable of penetrating clouds and atmospheric moisture, SAR provides all-weather, day-and-night observation capabilities\cite{cumming2005digital}, making it widely used in fields such as agriculture, forestry, geology, hydrology, disaster monitoring, and environmental surveillance\cite{osti_293027}. Current SAR research mainly falls into three directions: signal processing (e.g., motion compensation\cite{zhang2024iterative, chen2025data}, image reconstruction\cite{kang2023sar, wang2023atasi}, despeckling\cite{farhadiani2022sar, wang2025diffusion}, and maneuvering target imaging\cite{zhang2023pnp, jiang2025high}), computer vision (e.g., recognition\cite{hui2025bidirectional}, classification\cite{ni2024dpgunet}, and segmentation\cite{pena2024deepaqua}), and multimodal fusion with optical or hyperspectral data to enhance scene understanding\cite{deng2024hyperspectral, gu2025hpn}. This work primarily focuses on image reconstruction following motion compensation.

The core of SAR image formation lies in matched filtering (MF) in both range and azimuth\cite{cumming2005digital}, where classical algorithms such as the Range–Doppler method\cite{bamler2002comparison}, Chirp Scaling \cite{moreira2002extended}, $\omega$–K algorithm\cite{cumming2003interpretations}, and Back Projection\cite{yegulalp1999fast} coherently process the received echoes to produce focused images. Although effective in ideal conditions, these linear methods lack prior modeling and often suffer from sidelobe and grating-lobe artifacts under complex scenarios with closely spaced scatterers, strong interference, or incomplete and noisy data. The advent of compressed sensing (CS) \cite{candes2006robust, candes2006near, eldar2012compressed, eldar2015sampling} introduced sparsity priors into SAR imaging\cite{yang2013segmented, aberman2017sub, de2019compressed, hu2021compressive, kang2025compressive}, enabling high-quality reconstruction from limited measurements through algorithms such as FISTA\cite{beck2009fast, palomar2010convex}, ADMM\cite{boyd2011distributed}, OMP\cite{needell2009uniform, eldar2015sampling}, and SBL\cite{tipping2001sparse, wipf2004sparse}. While CS-based techniques effectively suppress sidelobes and enhance dominant scatterer reconstruction, their reliance on sparsity assumptions limits performance in scenes with distributed or textured targets, where richer prior information is needed to capture the underlying signal complexity.

In recent years, there has been research focused on improving measurement models to enhance the fidelity of imaging models \cite{xu2021image, dong2022high, bonfert2024improving}, as well as research focused on mining and utilizing richer prior information to better capture the inherent complex structures and statistical properties in real SAR scenes.
To characterize complex SAR scenes, researchers have explored combining or replacing sparsity with additional structural assumptions. An et al. exploited the two-dimensional correction structure of geosynchronous SAR scenes and incorporated low-rankness as a regularization constraint \cite{an2021geosynchronous, an2021joint}. Recognizing that strong scatterers violate low-rank assumptions yet still satisfy sparsity, they reformulated the problem as a joint sparse and low-rank recovery, further refining the signal model with a fourth-order slant range approximation. Bi et al. penalized group sparsity associated with azimuth ambiguities and ghost targets via the $\ell_{2,1}$-norm while enforcing overall sparsity with the $\ell_{1}$-norm \cite{bi2021sparse}. Fan et al. combined multiple priors, including $\ell_{2,1}$-norm constraints for azimuth ambiguity suppression, dictionary-based priors for feature enhancement, and TV/BM3D-based priors for speckle reduction, to simultaneously achieve low ambiguity, a high target-to-background ratio, and reduced speckle \cite{fan2024integrating}. While analytical priors such as sparsity or total variation have achieved remarkable progress, they often fail to capture the intricate structural and textural diversity of real SAR scenes.

In practice, SAR images exhibit highly heterogeneous backscatter patterns arising from multiple physical mechanisms and geometric configurations. The spatial distribution of these scatterers spans multiple scales, from fine object structures to broad terrain textures, resulting in nonstationary and scene-dependent statistics that are difficult to capture with fixed-form regularizers.

To address these challenges, recent studies have shifted toward learning-based, non-analytical priors that adaptively capture complex, scene-specific scattering behaviors and structural dependencies. Beyond directly modeling the signal domain, some approaches impose regularization within learned feature representations. For example, Yang \emph{et al.}~\cite{yang2024sar} replaced conventional sparsity priors with generalized, learnable image representations, achieving an end-to-end mapping from echoes to SAR images through deep unfolding. Similarly, Li \emph{et al.}~\cite{li2022stls} introduced a feature transformation operator into a deep unfolding framework, enabling adaptive prior learning while maintaining structured regularization constraints. By operating in a learned feature space, these methods enhance the expressive power of traditional priors and enable more flexible modeling of the complex, multiscale characteristics inherent in SAR imagery.

Inspired by the diverse and complex scattering characteristics of real SAR scenes, directly substituting the traditional regularization term with a learnable neural network offers an effective alternative. 
Embedding this network within an optimization-inspired unfolding architecture further mitigates the limitations of hand-crafted priors. 
Xiong et al. developed Lq-SPB-Net, a deep unfolding network that integrates nonconvex $\ell_q$-norm constraints with CNN-based projection operators, improving speckle suppression and edge preservation~\cite{xiong2021q}. 
Xu et al. leveraged the nonconvexity of the generalized $\ell_q$-norm to reduce bias and enhance target amplitudes, combining it with convolutional regularization to better preserve target structures~\cite{xu2024joint}. 
Wu et al. jointly optimized sparse SAR imaging and azimuth undersampling patterns, employing a learnable non-uniform sampling operator while using CNN-based regularization to suppress blur and noise~\cite{wu2024mf}. 
Li et al. proposed a hybrid CNN–self-attention regularizer capturing both local and global image features, effectively transforming inputs into a feature domain where targets and clutter are more easily distinguished~\cite{li2024sar}. 
By operating in a learnable manner, these deep priors flexibly model multi-scale, multi-directional, and scene-dependent scattering patterns that are difficult to capture with fixed structural priors. 
However, a key limitation remains that the network weights are tightly coupled to the specific measurement model used during training: changes in the sensing matrix or forward operator typically require retraining, limiting adaptability across different SAR acquisition scenarios.

To overcome the reliance of deep priors on fixed measurement operators, which limits adaptability across different SAR configurations, diffusion-based priors have recently emerged as a powerful alternative. As a class of generative models, diffusion priors capture intricate distributions of natural or SAR-specific backscatter patterns, achieving a better balance between prior knowledge and data fidelity through iterative, measurement-conditioned denoising. Moreover, diffusion-based inverse solvers provide a flexible means to incorporate complex image priors while maintaining adaptability to varying measurement models and waveform parameters.  Wang et al.~\cite{wang2024synthetic} introduced diffusion priors into SAR imaging and proposed a conditional-diffusion–based reconstruction method, demonstrating clear advantages over prior-free, hand-crafted prior, and CNN-based prior approaches. However, their method employs a guided scheme that approximates the posterior score using the prior score within the reverse diffusion process. This approximation ignores the nonlinear dependence between intermediate noisy states and the measurement data, leading to biased gradient updates, weakened measurement consistency, and residual artifacts such as sidelobes and texture over-smoothing. Notably, this approach can be viewed as an adaptation of Diffusion Posterior Sampling (DPS)~\cite{chung2022diffusion} to the SAR imaging setting.

To address these limitations, we pursue a tighter integration between measurement fidelity and prior modeling. Following the alternating sampling framework proposed by Xu \emph{et al.}~\cite{xu2024provably}, we alternate between likelihood- and prior-driven updates via proximal sampling. This progressive formulation preserves the expressive power of diffusion priors while rigorously enforcing measurement constraints, thereby mitigating the inherent bias of guided diffusion and yielding reconstructions with reduced artifacts and improved structural fidelity.

In this work, we extend the split Gibbs sampling architecture to SAR imaging, integrating diffusion priors with the physical measurement model to achieve measurement-consistent and high-fidelity reconstruction. This method effectively mitigates the limitations of standard diffusion approaches by maintaining consistency with the SAR forward model while exploiting high-order structural priors encoded in diffusion models. Extensive experiments on both simulated and real Sentinel-1A data validate the effectiveness of the method in SAR imaging. In simulation, this method achieves an average improvement of over 7~dB in PSNR, while the modified peak and integrated sidelobe ratios (MPLSR and MISLR) improve by at least 2.96~dB and 11.5~dB, respectively, compared with the best-performing baseline among matched filtering, compressed-sensing methods (e.g., FISTA, ADMM), and diffusion-prior-based approach (e.g., DPS). On real-world data, we observe an average PSNR gain of more than 1.6~dB. Fine structural features such as ridges and edges, often blurred or lost in competing methods, are clearly recovered, and noise as well as speckle are significantly suppressed, yielding sharper and more interpretable SAR images.

The rest of this paper is organized as follows. Section~\ref{background} introduces the background and problem formulation. Section~\ref{method} presents the Diffusion-Model-based inverse problem for SAR Imaging Reconstruction in detail. Section~\ref{experiments} reports experimental results. Section~\ref{conclusion} concludes the paper.

\section{Background and Problem Formulation}\label{background}
\subsection{Signal Model of SAR}

\begin{figure}[!t]
\centering
\includegraphics[width=3in]{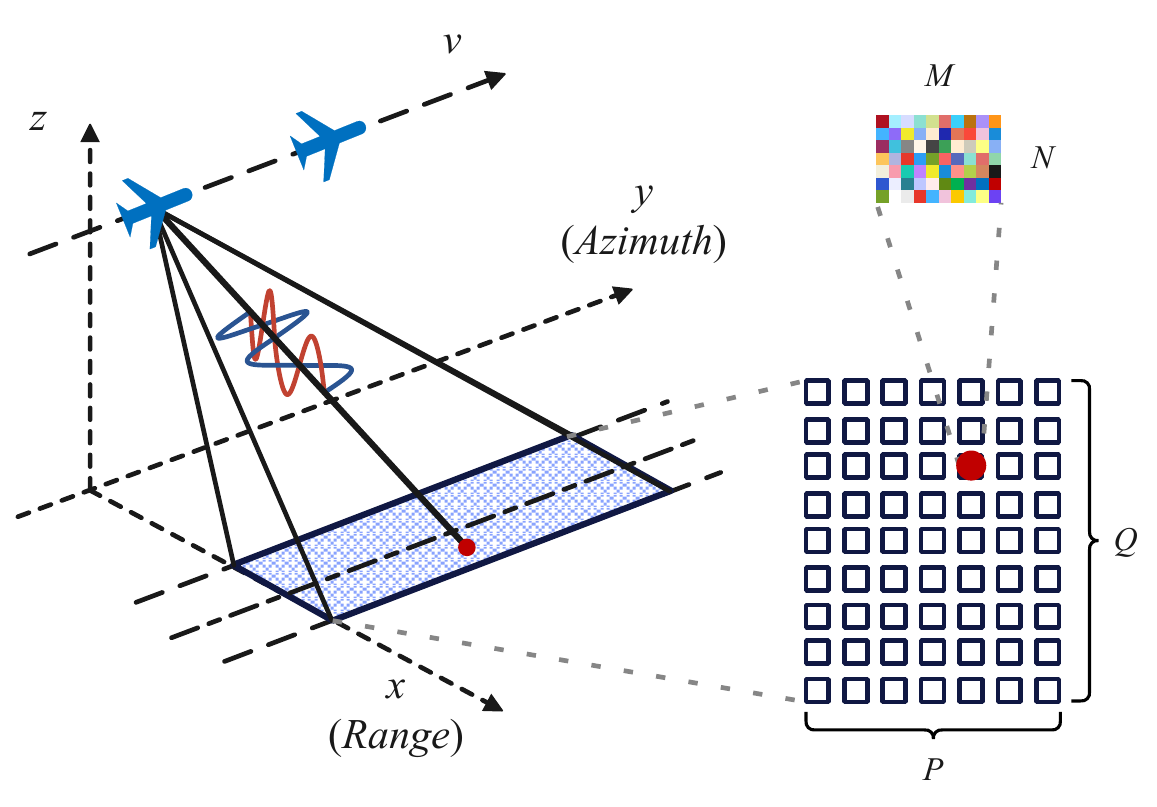}
\caption{SAR imaging geometry.}
\label{fig:sar_geo}
\end{figure}

As shown in Fig.~\ref{fig:sar_geo}, consider a SAR platform moving at a constant velocity $v$ along the $y$-axis while transmitting a linear frequency modulation (LFM) signal
\begin{equation}
s_{t} (\tau) = \exp{ \left( j 2 \pi f_{c} \tau \right)} \exp{ \left( j \pi K_{r} \tau^{2} \right)},
\end{equation}
where $\tau$ denotes the fast time, $f_{c}$ is the carrier frequency, and $K_{r}$ is the range modulation slope.

For a target located at $\boldsymbol{\omega} = (x, y)$ with backscattering coefficient $\beta_{\boldsymbol{\omega}}$, the received baseband echo is given by
\begin{equation}
\label{baseband_echo}
\begin{split}
    y_{\text{b}} ( \tau, \eta ) = & \, \beta_{\boldsymbol{\omega}}  \exp\!\left( - j \tfrac{4 \pi f_{c}}{c} R(\eta) \right) \\
    & \cdot \exp\!\left( j \pi K_{r}  \left( \tau - \tfrac{2 R(\eta)}{c} \right)^{2} \right),
\end{split}
\end{equation}
where $\eta$ denotes the slow time, $c$ is the speed of light, and $R(\eta)$ is the instantaneous distance between the target and radar, expressed as
\begin{equation}
\label{slant_range}
    R ( \eta ) = \sqrt{R_{x}^{2} + (v \eta - y )^{2}}
    \approx R_{x} + \frac{( v \eta - y )^{2}}{2R_{x}},
\end{equation}
with $R_{x}$ denoting the closest range between the radar and the target.
Under the stop-and-go assumption, the radar platform is considered stationary during each pulse duration, allowing us to approximate the echo as
\begin{equation}
\label{echo_simplify}
\begin{split}
    y_{\text{b}} ( \tau, \eta ) = &
    \beta_{\boldsymbol{\omega}} \exp\!\left( - j \tfrac{4 \pi f_{c} R_{x}}{c} \right) 
    \exp\!\left( - j \tfrac{2 \pi f_{c}}{R_{x}c} (v \eta - y)^{2} \right) \\
    & \cdot \exp\!\left( j \pi K_{r} \bigg( \tau - \tfrac{2 R_{x}}{c} \bigg)^{2} \right).
\end{split}
\end{equation}
Expanding \eqref{echo_simplify}, we obtain
\begin{equation}
\label{Signal decoupling}
\begin{split}
    y_{\text{b}} ( \tau, \eta ) = &
    \gamma_{\boldsymbol{\omega}} \exp\!\left( - j  \pi K_{a}  \eta^{2} \right)
    \exp\!\left( j 2 \pi \frac{K_{a}}{v} y \eta \right) \\
    & \cdot \exp\!\left( j  \pi K_{r}  \tau^{2} \right)
    \exp\!\left( - j 2 \pi \frac{ 2 K_{r}}{c} R_{x} \tau \right).
\end{split}
\end{equation}
where $\gamma_{\boldsymbol{\omega}} = \beta_{\boldsymbol{\omega}}  \exp\!\left( - j \tfrac{4 \pi f_{c} R_{x}}{c} - j \tfrac{2 \pi f_{c} y^{2}}{R_{x}c} + j 4 \pi K_{r} \tfrac{R_{x}^{2}}{c^{2}}\right)
$ and $K_{a} = \tfrac{2 f_{c} v^{2}}{c R_{x}}$ denote the azimuth modulation slope.

After dechirping in both range and azimuth, we obtain
\begin{equation}
\label{dechip}
    y_{\text{de}} ( \tau, \eta ) = 
    \gamma_{\boldsymbol{\omega}} \exp\!\left( j 2 \pi \tfrac{K_{a}}{v} y \eta \right)
    \exp\!\left( - j 2 \pi \tfrac{ 2 K_{r}}{c} R_{x} \tau \right).
\end{equation}
By sampling at fast time $\tau = n/f_{s}$ and slow time $\eta = m T_{r}$, we have
\begin{equation}
\label{fast_slow_sampling}
    y_{\text{de}}  [ n, m ] = 
    \gamma_{\boldsymbol{\omega}} \exp\!\left( j 2 \pi \tfrac{K_{a} T_{r}}{v} y  m \right)
    \exp\!\left( - j 2 \pi \tfrac{ 2 K_{r}}{c f_{s}} R_{x} n \right),
\end{equation}
where $T_{r}$ is the pulse repetition interval (PRI) and $f_{s}$ is the sampling frequency.  
Defining the range and azimuth frequencies as
\begin{equation}
\label{rg_az_frequency}
    f_{\text{rg}} = \tfrac{2 K_{r} }{c f_{s}} R_{x}, \quad
    f_{\text{az}} = \tfrac{K_{a} T_{r}}{v} y,
\end{equation}
we rewrite
\begin{equation}
    y_{\text{de}}  [ n, m ] = 
    \gamma_{\boldsymbol{\omega}} \exp( - j 2 \pi f_{\text{rg}} n )
    \exp( j 2 \pi f_{\text{az}} m ).
\end{equation}

The overall received signal is the superposition of echoes from all scatterers in the observed scene:
\begin{equation}
\label{final_received_signal}
    y [ n, m ] = 
    \iint \limits_{\Omega} \gamma_{\boldsymbol{\omega}} \exp( - j 2 \pi f_{\text{rg}} n )
    \exp( j 2 \pi f_{\text{az}} m )  \,\mathrm{d} \boldsymbol{\omega}.
\end{equation}
To facilitate processing, the scene is discretized into a $P \times Q$ grid, as shown in Fig.~\ref{fig:sar_geo}. Then, the integral in \eqref{final_received_signal} is approximated as
\begin{equation}
\label{discrete_received_signal}
    y [ n, m ] = 
    \sum_{p=0}^{P-1} \sum_{q=0}^{Q-1} \gamma_{p,q} 
    \exp( - j 2 \pi f_{\text{rg}}[p] n )
    \exp( j 2 \pi f_{\text{az}}[q] m ).
\end{equation}
The backscattering coefficients are represented by a complex-valued matrix
\begin{equation}
\label{backscattering_coefficient_martrix}
    \boldsymbol{X} =
    \begin{bmatrix}
         \gamma(p_{0}, q_{0}) & \cdots & \gamma(p_{0}, q_{Q-1}) \\
        \vdots & \ddots & \vdots \\
        \gamma(p_{P-1}, q_{0}) & \cdots & \gamma(p_{P-1}, q_{Q-1})
    \end{bmatrix} 
    \in \mathbb{C}^{P \times Q}.
\end{equation}
Defining the range steering vector $\boldsymbol{\varphi}_{p} \in \mathbb{C}^{N}$ and the azimuth steering vector $\boldsymbol{\psi}_{q} \in \mathbb{C}^{M}$:
\begin{equation}
\label{steering_vector}
\begin{split}
    \boldsymbol{\varphi}_{p}[n] = \exp( - j 2 \pi \tfrac{p}{P} n ), \,
    \boldsymbol{\psi}_{q}[m] = \exp( - j 2 \pi \tfrac{q}{Q} m ).
\end{split}
\end{equation}
The corresponding measurement matrices are
\begin{equation}
\label{measurement_matrix}
\begin{split}
    \boldsymbol{\Phi} &= [\boldsymbol{\varphi}_{0}, \cdots, \boldsymbol{\varphi}_{p}, \cdots, \boldsymbol{\varphi}_{P-1}] \in \mathbb{C}^{N \times P}, \\
    \boldsymbol{\Psi} &= [\boldsymbol{\psi}_{0}, \cdots, \boldsymbol{\psi}_{q}, \cdots, \boldsymbol{\psi}_{Q-1}] \in \mathbb{C}^{M \times Q}.
\end{split}
\end{equation}
Both $\boldsymbol{\Phi}$ and $\boldsymbol{\Psi}$ are expressed as row-wise Fourier matrices:
\begin{equation}
\label{row_wise_Fourier_matrices}
\boldsymbol{\Phi} = \boldsymbol{\Pi}_{r} \boldsymbol{F}_{r}, \quad
\boldsymbol{\Psi} = \boldsymbol{\Pi}_{a} \boldsymbol{F}_{a},
\end{equation}
where $\boldsymbol{\Pi}_{r}$ and $\boldsymbol{\Pi}_{a}$ are row-selection matrices.
Thus, the SAR signal model in \eqref{discrete_received_signal} can be expressed in matrix form as
\begin{equation}
\label{matrix_model}
    \boldsymbol{Y}  =  \boldsymbol{\Phi} \boldsymbol{X} \boldsymbol{\Psi}^{\mathrm{H}} + \boldsymbol{W},
\end{equation}
where $\boldsymbol{W} \in \mathbb{C}^{N \times M}$ denotes Gaussian noise.  
Equivalently, the model is written in vectorized form:
\begin{equation}
\label{vector_model}
    \boldsymbol{y}  =  \boldsymbol{A} \boldsymbol{x} + \boldsymbol{w},
\end{equation}
where $\boldsymbol{y} = \mathrm{vec}(\boldsymbol{Y}) \in \mathbb{C}^{NM}$, $\boldsymbol{x} = \mathrm{vec}(\boldsymbol{X}) \in \mathbb{C}^{PQ}$, $\boldsymbol{A} = \boldsymbol{\Psi}^{\ast} \otimes \boldsymbol{\Phi}  \in \mathbb{C}^{NM \times PQ}$, $\otimes$ denotes the Kronecker product, and $\boldsymbol{w} \sim \mathcal{CN} ( \boldsymbol{0}, \sigma^{2}\boldsymbol{I} )$.

Under the measurement model \eqref{vector_model}, the goal is to estimate the target scene $\boldsymbol{x}$ from the received echoes $\boldsymbol{y}$, given the known measurement matrix $\boldsymbol{A}$.

\subsection{Challenges of Matched Filter-based SAR Imaging with Incomplete Measurements}

Ideally, in real SAR data processing, the dimensions of the measurement and scene are balanced, i.e., $NM = PQ$ \cite{8344566}. 
Under this condition, the backscattering coefficients $\boldsymbol{x}$ can be estimated through matched filtering as
\begin{equation}
  \hat{\boldsymbol{x}} = \boldsymbol{A}^{\mathrm{H}} \boldsymbol{y},
\end{equation}
where $(\cdot)^{\mathrm{H}}$ denotes the Hermitian transpose. 
In the ideal case where $\boldsymbol{\Phi} = \boldsymbol{F}_{r}$ and $\boldsymbol{\Psi} = \boldsymbol{F}_{a}$, $\boldsymbol{A}$ is unitary, i.e., $\boldsymbol{A}^{\mathrm{H}} \boldsymbol{A} = \boldsymbol{I}$. 
Consequently, matched filtering provides an unbiased estimate of $\boldsymbol{x}$,
\begin{equation}
  \mathbb{E}[\hat{\boldsymbol{x}}] = \boldsymbol{x},
\end{equation}
so $\hat{\boldsymbol{x}}$ can be considered a recovery of $\boldsymbol{x}$, 
although residual noise remain in practice.

However, in practical SAR systems, the received echo data $\boldsymbol{y}$ are often \emph{incomplete} and noisy due to several practical limitations.
On the one hand, certain portions of the received signals may be affected by interference or jamming \cite{dong2014sar}, which must be removed to avoid degrading the reconstruction quality.
On the other hand, transmitting the full-resolution echo data imposes a heavy communication and storage burden, especially for high-resolution SAR systems covering large areas, so subsampling or selective acquisition is commonly employed \cite{kelly2012advanced, patel2010compressed}.
Consequently, the equivalent linear system in~\eqref{vector_model} becomes underdetermined, i.e., $NM < PQ$. 
Under such conditions, infinitely many $\boldsymbol{x}$ can satisfy $\boldsymbol{y} = \boldsymbol{A}\boldsymbol{x}$, making the reconstruction problem ill-posed. 
Directly applying matched filtering in this case yields not only an approximate estimate of $\boldsymbol{x}$ but also inevitably introduces strong sidelobes and artifacts, since data incompleteness violates the unitary property of the measurement matrix.

Therefore, the SAR imaging process is naturally formulated as an inverse problem, 
where additional prior knowledge is incorporated to assist reconstruction. 
A well-defined prior distribution provides strong guidance when the measurement data are incomplete and noisy, resulting in stable and high-fidelity estimates of the reconstructed image.

\section{SAR Imaging with Prior Knowledge}

\subsection{Maximum a Posteriori Estimation Framework}
To achieve a stable and physically meaningful reconstruction, it is essential to introduce additional \emph{prior information} about $\boldsymbol{x}$ to regularize the solution space. One common approach is to formulate the reconstruction as a \emph{maximum a posteriori} (MAP) estimation:
\begin{equation}
\label{eq:inverse_problem}
\begin{split}
  \hat{\boldsymbol{x}}
  &= \mathop{\arg\max}_{\boldsymbol{x}}\, p(\boldsymbol{x} \vert \boldsymbol{y}) \\
  &= \mathop{\arg\max}_{\boldsymbol{x}}\, p(\boldsymbol{y}\vert \boldsymbol{x})\,p(\boldsymbol{x}) \\
  &= \mathop{\arg\min}_{\boldsymbol{x}}\,
     \mathcal{L}(\boldsymbol{x};\boldsymbol{y}) + f(\boldsymbol{x}),
\end{split}
\end{equation}
where $\mathcal{L}(\boldsymbol{x};\boldsymbol{y})=-\log p(\boldsymbol{y} \vert \boldsymbol{x})$ is the likelihood potential (data fidelity term), and $f(\boldsymbol{x})=-\log p(\boldsymbol{x})$ denotes the prior potential (regularization term).
Common choices for analytical priors include sparsity-promoting regularizers such as the $\ell_1$-norm, $f(\boldsymbol{x}) = \mu \Vert\boldsymbol{x}\Vert_1$, and smoothness-promoting regularizers such as Total Variation (TV), $f(\boldsymbol{x}) = \mu \Vert\boldsymbol{x}\Vert_{\text{TV}}$ \cite{palomar2010convex}. The parameter $\mu > 0$ controls the strength of the regularization.

While effective in some scenarios, these analytical priors often fail to capture the rich, multi-scale, and directionally varying structures found in real-world data. By oversimplifying the true statistics of $\boldsymbol{x}$, they frequently lead to over-smoothing and the loss of fine details~\cite{li2024sar}. This is particularly problematic in applications like SAR, where scenes exhibit highly diverse and complex scattering behaviors. For example, in mountainous regions, the terrain produces continuous ridges and valleys, high-contrast edges, and spatially correlated speckle patterns, leading to textured and highly structured backscatter that is challenging to model with simple priors.

To overcome these limitations, it is desirable to employ more expressive data-driven priors, denoted as $p_{\theta}(\boldsymbol{x})$, 
where $\theta$ represents the learnable parameters of the prior model. 
These priors are learned from a dataset $\mathcal{X}$ of SAR images by neural network, allowing them to better capture the high-dimensional distribution of real SAR scenes. 
However, since such priors typically lack explicit analytical forms, their direct integration into optimization-based MAP frameworks remains a significant challenge.

\subsection{Posterior Sampling-Based MAP Realization}
For problems involving an analytical likelihood but non-analytical priors, \emph{posterior sampling} is a powerful solution. It provides a flexible framework that effectively and "softly" achieves MAP estimation where direct analytical methods fail. Instead of explicitly solving \eqref{eq:inverse_problem}, one samples from the posterior distribution:
\begin{equation}
\label{eq:posterior_sampling}
\begin{split}
  \hat{\boldsymbol{x}} &\sim p(\boldsymbol{x} \vert \boldsymbol{y}) 
  \propto p(\boldsymbol{y} \vert \boldsymbol{x})\,p_{\theta}(\boldsymbol{x}) \\
  &= \exp\!\left\{-\mathcal{L}(\boldsymbol{x};\boldsymbol{y}) + \log p_{\theta}(\boldsymbol{x})\right\}.
\end{split}
\end{equation}

Typical Markov Chain Monte Carlo (MCMC) sampling methods, such as Langevin Dynamics (LD) \cite{welling2011bayesian} and the Metropolis-Hastings (MH) \cite{hastings1970monte} algorithm, are widely used for posterior inference. However, they exhibit significant differences in their informational requirements, which directly impacts their computational feasibility, particularly when dealing with complex, deep learning-based distributions.

The LD\cite{welling2011bayesian} is a gradient-based sampling method. Its core principle is to simulate a particle's movement, subject to random noise, across the potential energy surface defined by the target log-density. It utilizes the gradient of the log-density (the \emph{score function}) to drive the sample towards regions of higher probability. The LD update equation is given by:
\begin{equation}
\label{eq:ld_sampling}
    \boldsymbol{x}^{\star} = \boldsymbol{x} + \epsilon \nabla \log p(\boldsymbol{x} \vert \boldsymbol{y}) + \sqrt{2 \epsilon} \boldsymbol{w}.
\end{equation}
Here, the score function $\nabla \log p(\boldsymbol{x} \vert \boldsymbol{y}) = -\nabla \mathcal{L}(\boldsymbol{x};\boldsymbol{y}) + \nabla \log p_{\theta}(\boldsymbol{x})$, where $\nabla \log p_{\theta}(\boldsymbol{x})$ is directly learned from data using techniques such as denoising score matching \cite{song2019generative}, $\epsilon$ denotes the step size, and $\boldsymbol{w}$ is the injected Gaussian noise. The key advantage of LD is that it bypasses the need for explicit density function evaluation, relying only on its gradient, thus making it inherently advantageous for models with implicit density definitions.

MH algorithm~\cite{hastings1970monte} generates samples from a target distribution by iteratively performing a random exploration and acceptance decision. 
At each iteration, a candidate state $\boldsymbol{x}^{\star}$ is drawn from a proposal distribution $q(\boldsymbol{x}^{\star} \vert \boldsymbol{x})$, representing a stochastic search around the current point. 
This candidate is then accepted or rejected according to an acceptance criterion that compares how well $\boldsymbol{x}^{\star}$ matches the desired target distribution relative to the current state. 

The MH sampling requires evaluating the density ratio $p_{\theta}(\boldsymbol{x}^{\star}) / p_{\theta}(\boldsymbol{x})$, which is generally intractable for complex neural network–based priors.
Moreover, the simplicity of its proposal distribution often causes random-walk behavior and low sampling efficiency, particularly in high-dimensional spaces.
In contrast, LD sampling leverages the gradient of the log-probability to guide updates toward regions of higher posterior density, thereby improving convergence efficiency and exploration stability.

Beyond selecting an appropriate sampling strategy, learning a well-structured prior distribution is equally critical for effective posterior inference. 
While, in principle, the gradient of the data distribution can be estimated via denoising score matching, the remarkable capability of diffusion models to represent complex, high-dimensional data makes them a more practical and expressive choice. 
Notably, from the perspective of stochastic differential equations (SDEs), these two approaches are fundamentally unified \cite{song2020score}. 
In SAR imaging, such priors are typically learned from a dataset $\mathcal{X}$ of historical SAR images, providing a powerful data-driven prior for posterior sampling in inverse problems.

\section{Diffusion Models as Learnable Priors}
Diffusion models~\cite{sohl2015deep} provide a powerful and conceptually clear framework for solving inverse problems with data-driven priors.They learn the score function of the data distribution and thus implicitly capture fine-scale statistical structures of SAR scenes.  The learned prior is embedded into posterior sampling schemes, enabling a principled fusion of physics-based measurement models and data-driven priors, leading to high-fidelity and artifact-suppressed SAR reconstructions.
Here, diffusion models offer a flexible way to incorporate powerful learned priors into inverse problems, but in most current applications, they serve primarily as generative samplers rather than prior models.  

The central idea consists of two key components: \emph{prior learning} and \emph{prior utilization}. 
\textbf{Prior learning} aims to capture the underlying distribution of clean SAR images from training data, enabling the diffusion model to represent realistic spatial correlations, multi-scale structures, and the intrinsic statistics of SAR scenes. 
Unlike hand-crafted regularizations, diffusion priors are data-driven and can adaptively model complex, scene-dependent characteristics that are difficult to express analytically. 
\textbf{Prior utilization}, on the other hand, directly employs the pretrained diffusion model without further retraining. 
During reconstruction, the learned prior is integrated with the physical measurement model through posterior sampling, allowing each iterative update to move toward regions of higher posterior probability, simultaneously enforcing data fidelity and prior consistency.

\subsection{Prior Learning with Diffusion Model} \label{Prior Learning}
\begin{figure}[!t]
\centering
\includegraphics[width=3.5in]{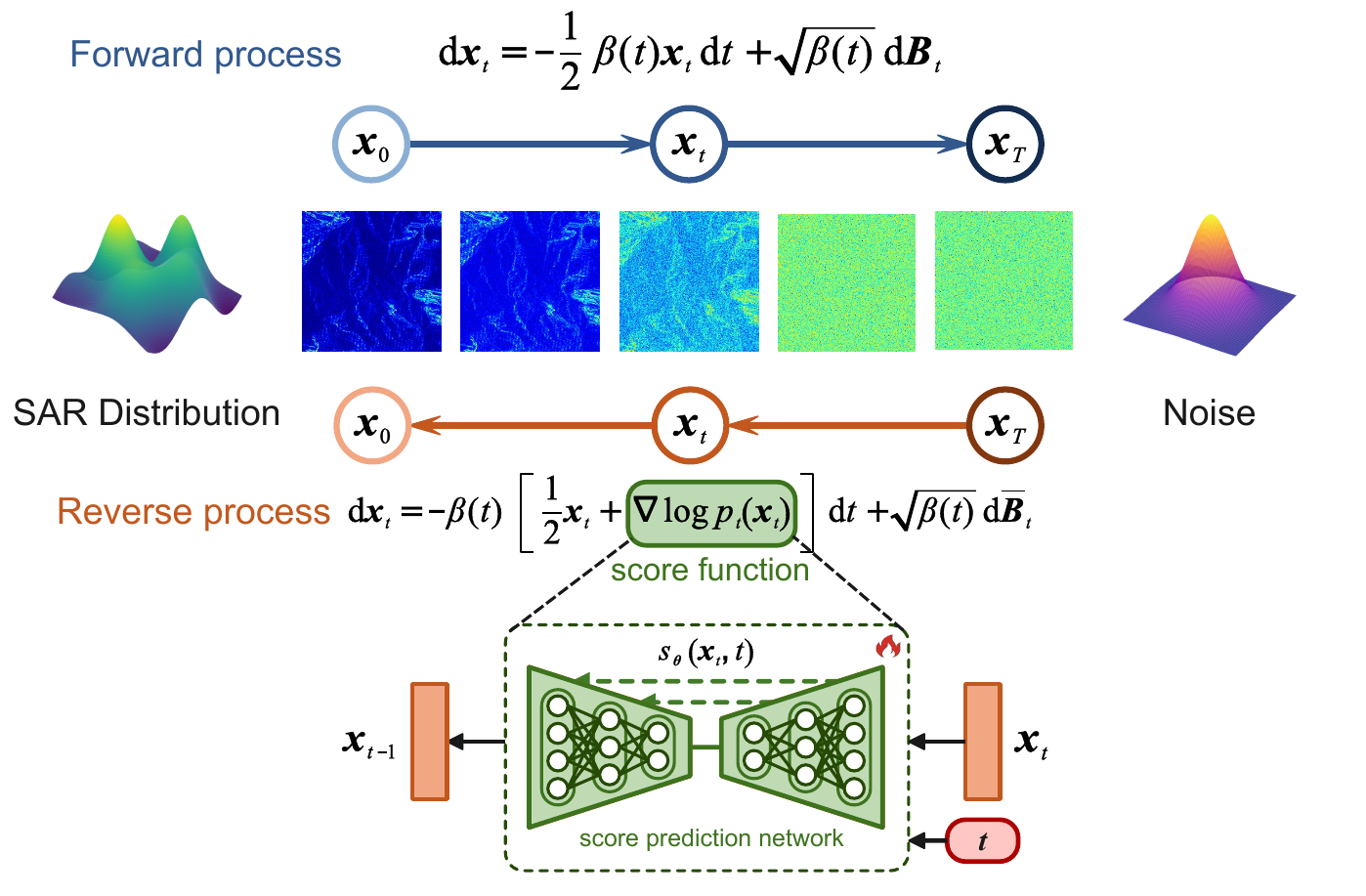}
\caption{SAR prior learning with diffusion model.}
\label{fig:prior_learning}
\end{figure}
As shown in Fig.~\ref{fig:prior_learning}, in the prior learning stage, diffusion models represent the SAR prior distribution  $p(\boldsymbol{x}_0)$ through a stochastic diffusion denoising process. 
Clean SAR images $\boldsymbol{x}_0 \sim p(\boldsymbol{x}_0)$ are gradually corrupted with Gaussian noise, producing a sequence of noisy states $\boldsymbol{x}_t$ described by the forward SDE:
\begin{equation}
\label{eq:forward_sde}
    d \boldsymbol{x}_t = -\tfrac{1}{2}\beta(t)\boldsymbol{x}_t\,dt + \sqrt{\beta(t)}\,d \boldsymbol{B}_t,
\end{equation}
where $\boldsymbol{B}_t$ denotes standard Brownian motion, and $\beta(t)$ controls the noise level over time. 
As $t$ increases, the SAR images approach pure Gaussian noise. 
The model is then trained to reverse this process via the reverse-time SDE:
\begin{equation}
\label{eq:reverse_sde}
    d \boldsymbol{x}_t = -\beta(t)\!\left[\tfrac{1}{2}\boldsymbol{x}_t + \nabla \log p_t(\boldsymbol{x}_t)\right] dt + \sqrt{\beta(t)}\, d \bar{\boldsymbol{B}}_t,
\end{equation}
which gradually denoises $\boldsymbol{x}_t$ back toward the SAR data manifold. 
In practice, a neural network $\boldsymbol{s}_\theta(\boldsymbol{x}_t,t)$ is trained to approximate the score function $\nabla \log p_t(\boldsymbol{x}_t)$ by denoising score matching:
\begin{equation}
\label{train_object}
    \min_{\boldsymbol{\theta}} \mathbb{E}_{t,p_{\boldsymbol{x}_t},p_0}
    \!\left[ \tfrac{1}{2}\! \left\Vert 
    \boldsymbol{s}_{\boldsymbol{\theta}}(\boldsymbol{x}_t,t)
    - \nabla \log p_{\boldsymbol{x}_t}(\boldsymbol{x}_t \vert \boldsymbol{x}_0)
    \right\Vert^2 \right],
\end{equation}
and in most implementations, the network predicts the noise instead of the score, which is equivalently converted as \cite{luo2022understanding}:
\begin{equation}
\label{noise_score}
    \boldsymbol{s}_{\boldsymbol{\theta}}(\boldsymbol{x}_t,t)
    = -\frac{1}{\sqrt{1 - e^{-2t}}}\boldsymbol{\epsilon}_{\boldsymbol{\theta}}(\boldsymbol{x}_t,t).
\end{equation}
Once trained, this score function effectively encodes the geometry of the SAR data manifold in a high-dimensional space, shaped by both physical and structural regularities.

For practical implementation, and to ensure compatibility with standard neural network architectures, the complex-valued variables are represented in a two-channel real-valued form, corresponding to their real and imaginary components.

Although diffusion models provide a general and powerful framework for prior representation, their application in SAR imaging remains relatively limited. Here, we employ diffusion models to capture the complex statistical characteristics of SAR scenes, such as structured spatial correlations of ridges and valleys, high-contrast edges, and terrain-specific speckle patterns in mountainous regions. Moreover, we found that even when diffusion models are adopted as priors, the specific manner in which the prior is incorporated into the reconstruction process can lead to significant differences in performance. This motivates the development of an effective utilization strategy tailored for SAR imaging.

\subsection{Guidance Method for Prior Utilization}
In order to efficiently utilize the learned diffusion prior in the posterior sampling of SAR imaging, a  straightforward approach is to modify the reverse SDE by replacing the unconditional score with the conditional score:
\begin{equation}
\nabla \log p(\boldsymbol{x}_t \vert\boldsymbol{y})
= \nabla \log p_t(\boldsymbol{x}_t)
+\nabla \log p(\boldsymbol{y} \vert \boldsymbol{x}_t),
\end{equation}
where the first term is provided by the learned diffusion score $\boldsymbol{s}_\theta(\boldsymbol{x}_t,t)$ and the second term corresponds to the likelihood gradient. 
However, the likelihood term is generally intractable because the measurement $\boldsymbol{y}$ depends only on the clean signal $\boldsymbol{x}_0$, not on its diffused counterpart $\boldsymbol{x}_t$. 

To overcome this problem, Diffusion Posterior Sampling (DPS)\cite{chung2022diffusion}, a typical guidance method, introduces $\boldsymbol{x}_0$ as a latent bridge and approximates:
\begin{equation}
\nabla \log p(\boldsymbol{y} \vert \boldsymbol{x}_t)
\approx \nabla \log p(\boldsymbol{y} \vert \hat{\boldsymbol{x}}_0),
\end{equation}
where $\hat{\boldsymbol{x}}_0 = \mathbb{E}[\boldsymbol{x}_0 \vert \boldsymbol{x}_t]$ is computed using Tweedie’s formula under a Gaussian assumption. 
This approximation makes posterior sampling feasible but introduces a fundamental bias: 
the single posterior mean $\hat{\boldsymbol{x}}_0$ cannot fully represent the nonlinear dependence between $\boldsymbol{y}$ and $\boldsymbol{x}_t$. 
Consequently, the approximated likelihood gradient no longer corresponds to the true $\nabla \log p(\boldsymbol{y} \vert \boldsymbol{x}_t)$, causing the diffusion trajectory to deviate from the learned data manifold. 

This bias alters the fundamental cost \eqref{eq:inverse_problem} and results in artifacts, oversmoothing, or reduced fidelity in the reconstructed SAR images.
Therefore, guidance-based diffusion solvers such as DPS are regarded as \emph{approximate guidance methods} rather than exact posterior samplers, they steer the generation toward measurement-consistent regions but do not rigorously follow the true posterior distribution.

\subsection{Asymptotic Method for Prior Utilization}
In order to sample from the true posterior distribution, asymptotic sampling methods converge to the target distribution as the number of iterations tends to infinity.  
The \emph{Split Gibbs Sampler (SGS)}~\cite{vono2019split, 10541919} facilitates this convergence by decoupling the data fidelity term and the prior term while maintaining consistency between them.  
Specifically, SGS introduces an auxiliary variable $\boldsymbol{z}$ and defines the following augmented joint distribution:
\begin{equation}
\label{SGS}
q(\boldsymbol{x}, \boldsymbol{z}) \propto 
\exp \Big( 
    - \mathcal{L}(\boldsymbol{z};\boldsymbol{y}) 
    - f_{\theta}(\boldsymbol{x})
    - \frac{1}{2\lambda^{2}} \Vert \boldsymbol{x} - \boldsymbol{z} \Vert^{2} 
\Big),
\end{equation}
where $\mathcal{L}(\boldsymbol{z};\boldsymbol{y})$ denotes the data fidelity potential, $f_{\theta}(\boldsymbol{x})$ represents the learned prior potential, and $\lambda$ controls the coupling strength between $\boldsymbol{x}$ and $\boldsymbol{z}$.  
The quadratic term acts as a soft consistency constraint that ties the two variables while still allowing stochastic flexibility.

Although direct sampling from \eqref{SGS} is intractable, its structure naturally leads to two conditional distributions:
\begin{equation}
\label{cond_z}
q(\boldsymbol{z} \vert \boldsymbol{x}) \propto
\exp \left( 
    - \mathcal{L}(\boldsymbol{z};\boldsymbol{y}) 
    - \frac{1}{2\lambda^2} \Vert \boldsymbol{x} - \boldsymbol{z}\Vert ^2 
\right),
\end{equation}
\begin{equation}
\label{cond_x}
q(\boldsymbol{x} \vert \boldsymbol{z}) \propto
\exp \left( 
    - f_{\theta}(\boldsymbol{x}) 
    - \frac{1}{2\lambda^2} \Vert \boldsymbol{x} - \boldsymbol{z}\Vert ^2 
\right).
\end{equation}
This decomposition allows iterative sampling from the conditional distributions, forming a Markov chain whose stationary distribution asymptotically approaches the true posterior. 

Xu et al. \cite{xu2024provably} provided a specific strategy of likelihood sampling and prior sampling. Specifically, the likelihood term is sampled using Langevin dynamics:
\begin{equation}
\label{eq:LD_SDE}
    d\boldsymbol{z}_{k}^{(t)} 
    = \left[- \nabla \mathcal{L}\big(\boldsymbol{z}_{k}^{(t)}; \boldsymbol{y}\big)
    - \tfrac{1}{\lambda^{2}}\big(\boldsymbol{z}_{k}^{(t)} - \boldsymbol{x}_{k}\big) \right] dt 
    + \sqrt{2}\, d\boldsymbol{B}^{(t)},
\end{equation}
where $\boldsymbol{B}^{(t)}$ denotes a standard Brownian motion.
The exponential integrator (EI)\cite{zhang2022fast} is applied to solve \eqref{eq:LD_SDE}. Let $\kappa$ denote the Langevin step size; the update of $\boldsymbol{z}$ over each interval $[t\kappa, (t+1)\kappa)$ is expressed as
\begin{equation}
\label{Langevin_Dynamics_EI}
\begin{split}
\boldsymbol{z}_{k}^{(t+1)} 
= & \; e^{-\kappa / \lambda^{2}}\, \boldsymbol{z}_{k}^{(t)} 
+ \big(1 - e^{-\kappa / \lambda^{2}}\big)\, \boldsymbol{x}_{k} \\
& - \lambda^{2}\big(1 - e^{-\kappa / \lambda^{2}}\big)\,
\nabla \mathcal{L}\big(\boldsymbol{z}_{k}^{(t)}; \boldsymbol{y}\big) \\
& + \lambda \sqrt{1 - e^{-2\kappa / \lambda^{2}}}\, \boldsymbol{w}^{(t)},
\end{split}
\end{equation}
where $\boldsymbol{w}^{(t)}$ is Gaussian noise.

However, for the prior term, since the implicit prior learned by a diffusion model does not admit an explicit analytical density, direct sampling from this distribution $q(\boldsymbol{x} \vert \boldsymbol{z}_{k}) $ is intractable. 
Recall from \eqref{cond_x} that, according to Bayes’ rule, the conditional distribution is expressed as 
$q(\boldsymbol{x} \vert \boldsymbol{z}_{k}) \propto p(\boldsymbol{z}_{k} \vert \boldsymbol{x})\, p_{\theta}(\boldsymbol{x})$.
This formulation naturally leads to a noisy model:
\begin{equation}
\label{noisy_model}
    \boldsymbol{z}_{k} = \boldsymbol{x} + \boldsymbol{w}, \quad 
    \boldsymbol{w} \sim \mathcal{N}(\mathbf{0}, \lambda^{2}\mathbf{I}).
\end{equation}
According to the relationship established in Equation~\eqref{noisy_model},  the deterministic Denoising Diffusion Implicit Model (DDIM)~\cite{song2020denoising}  is employed as a prior sampler.  Specifically, let $\boldsymbol{\xi}$ denote a random variable satisfying $p_{\boldsymbol{\xi}}(x) \sim p(\boldsymbol{x}=x \vert \boldsymbol{z}_{k})$.  Define the residual variable $\boldsymbol{\zeta} = \boldsymbol{\xi} - \boldsymbol{z}_{k}$, and construct an Ornstein–Uhlenbeck (OU) process for $\boldsymbol{\zeta}^{(t)}$ as
\begin{equation}
\label{ou_zeta}
    d\boldsymbol{\zeta}^{(t)} = -\boldsymbol{\zeta}^{(t)}\, dt 
    + \sqrt{2}\, d\boldsymbol{B}^{(t)}, 
    \quad \boldsymbol{\zeta}^{(0)} \sim p_{\boldsymbol{\zeta}},
\end{equation}
where $\boldsymbol{B}^{(t)}$ denotes standard Brownian motion.
The corresponding probability flow ODE is then given by
\begin{equation}
\label{ode_zeta}
    d\boldsymbol{\zeta}^{(t)} 
    = -\left[\boldsymbol{\zeta}^{(t)} 
    + \nabla \log p_{\boldsymbol{\zeta}^{(t)}}(\boldsymbol{\zeta}^{(t)})\right] dt.
\end{equation}
According to the connection between $p_{\boldsymbol{\zeta}^{(t)}}$ and $p_{\boldsymbol{x}_{t}}$\cite{xu2024provably}, the score function is expressed as
\begin{equation}
\label{score_final}
\begin{split}
    \nabla \log p_{\boldsymbol{\zeta}^{(t)}} (\boldsymbol{\zeta}^{(t)}) 
    = & - \frac{e^{2t} \boldsymbol{\zeta}^{(t)}}{\lambda^{2} + e^{2t} -1}  
    + \frac{ e^{t-\tau} \lambda^{2}}{\lambda^{2} + e^{2t} -1} \, \\
    & \cdot \boldsymbol{s}_{\boldsymbol{\theta}} \!\left(
      e^{-\tau} \boldsymbol{z}_{k} 
      + \frac{ e^{t-\tau} \lambda^{2} \boldsymbol{\zeta}^{(t)}}{\lambda^{2} + e^{2t} -1}, \tau 
    \right),
\end{split}
\end{equation}
where $\tau = \frac{1}{2}\log\!\left(\frac{\lambda^{2}(e^{2t}-1)}{\lambda^{2}+e^{2t}-1}+1\right)$.

Discretizing \eqref{ode_zeta} with an EI yields the deterministic DDIM-style update:
\begin{equation}
\label{deterministic_sample}
\begin{split}
    \boldsymbol{\zeta}^{(i-1)} =& \tfrac{\sqrt{(\lambda^{2}-1) r_{i-1} + 1}}{\sqrt{(\lambda^{2}-1) r_{i} + 1}} \boldsymbol{\zeta}^{(i)} + \sqrt{(\lambda^{2}-1) r_{i-1} + 1} \\
     & \cdot \left( \arctan \tfrac{\lambda}{\sqrt{r_{i}^{-1}-1}} - \arctan \tfrac{\lambda}{\sqrt{r_{i-1}^{-1}-1}} \right) \\
    & \cdot \boldsymbol{\epsilon}_{\boldsymbol{\theta}} \!\left( \sqrt{\bar{\alpha}_{i}} \boldsymbol{z}_{k} + \tfrac{ \lambda^{2}  \sqrt{r_{i} \bar{\alpha}_{i}} \boldsymbol{\zeta}^{(i)}}{ (\lambda^{2} - 1)r_{i} + 1} \right),
\end{split}
\end{equation}
where $r_i = \frac{(\lambda^{2}+1) \bar{\alpha}_i - 1}{\lambda^{2} + \bar{\alpha}_i - 1}$ and $\boldsymbol{\epsilon}_{\boldsymbol{\theta}}$ and $\boldsymbol{s}_{\boldsymbol{\theta}}$ are related by \eqref{noise_score}.  
Finally, the denoised estimate is obtained as
\[
\boldsymbol{x}_{k} = \boldsymbol{z}_{k} + \boldsymbol{\zeta}^{(0)}.
\]

\subsection{Discussion}
Guidance methods directly approximate the likelihood score to adapt the diffusion process for inverse sampling, thereby steering samples toward consistency with measurements. This approach is computationally efficient and easy to implement, but it alters the original objective, resulting in only an approximation of the true posterior distribution. In contrast, asymptotic methods construct a Markov chain whose stationary distribution asymptotically converges to the true posterior. Although they generally require more iterations to reach convergence, they preserve the original posterior objective, offering better physical consistency and reconstruction fidelity. This property is particularly advantageous in SAR imaging, where the measurement model is underdetermined and noisy, and the underlying prior distribution is often complex, as it allows progressive methods to more faithfully characterize uncertainty and maintain physically consistent reconstructions.
For underdetermined and noisy SAR imaging problems, asymptotic methods are generally preferred.

\section{Implementation of Diffusion-based Posterior Split Gibbs Sampling for SAR Imaging} \label{method}
Building upon the method of Xu et al.~\cite{xu2024provably}, we adapt it to the SAR imaging setting with incomplete measurements. Unlike typical image-reconstruction tasks, which involve single-channel or RGB real-valued data, SAR signals are inherently complex-valued, requiring several implementation refinements. 

Our approach is structured around two key components: \emph{prior learning} and \emph{prior utilization}. In the prior learning stage, the diffusion model is trained on complex-valued SAR images represented as two real-valued channels. This formulation enables the network to effectively capture SAR-specific backscatter statistics, multi-scale structures, and intrinsic spatial correlations. During the prior utilization stage, the pretrained diffusion prior is incorporated into posterior sampling alongside the SAR forward model. Likelihood sampling is performed directly in the complex domain, while prior sampling is conducted in the two-channel representation, ensuring both physical consistency and compatibility with standard neural network modules.

\subsection{Prior Learning for SAR}
SAR images are inherently complex-valued, so each image is converted into a two-channel real-valued representation, with the real and imaginary components treated as separate channels. This allows the diffusion model to operate on standard real-valued neural network architectures while preserving the information content of complex SAR signals.

\begin{figure*}[!t]
\centering
\includegraphics[width=6.5in]{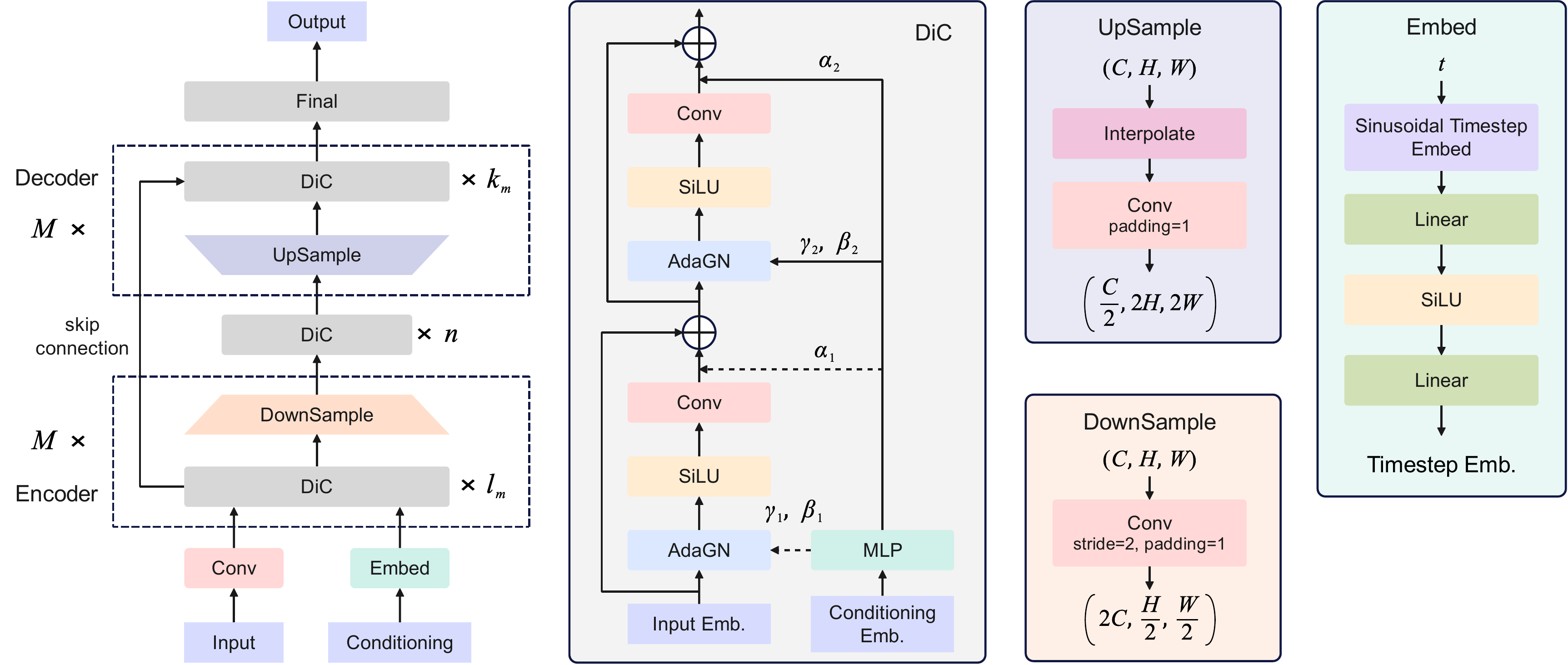}
\caption{Architecture of the noise prediction network $\boldsymbol{\epsilon}_{\boldsymbol{\theta}}$.}
\label{Noise_prediction}
\end{figure*}

As shown in Fig.~\ref{Noise_prediction}, the noise prediction network $\boldsymbol{\epsilon}_{\boldsymbol{\theta}}$ employed in our diffusion model adopts a U-shaped architecture constructed from Diffusion CNN (DiC) blocks~\cite{tian2025dic}. The U-Net consists of encoder and decoder pathways connected via a middle layer, while skip connections bridge corresponding encoder–decoder pairs to facilitate information flow and alleviate gradient vanishing issues.
Each DiC block contains two consecutive convolutional layers, where each layer sequentially applies Group Normalization (GN), the SiLU activation, and a $3\times3$ convolution. Conditional information, such as the diffusion timestep~$t$, is incorporated via Adaptive Group Normalization (AdaGN), formulated as:
\begin{equation}
\label{AdaGN}
\begin{split}
    & \gamma, \beta, \alpha = \text{MLP}\!\left(\text{Embed}(t) \right), \\
    & \boldsymbol{X}^{\prime} = \text{AdaGN}(\boldsymbol{X}) = \gamma \cdot \text{GN}(\boldsymbol{X}) + \beta, \\
    & \boldsymbol{X}^{\prime\prime} = \boldsymbol{X} + \alpha \cdot \text{Conv}\!\left(\text{SiLU}\!\left(\boldsymbol{X}^{\prime}\right)\right).
\end{split}
\end{equation}
In practice, the second convolutional layer in each DiC block always receives the conditional input, whereas the first layer is optionally conditioned depending on the dataset and task.

With the network architecture defined, the diffusion model is trained to predict the additive Gaussian noise at each diffusion timestep. Specifically, given a clean SAR sample $\boldsymbol{x}_0 \in \mathcal{X}$ and a diffusion timestep $t$, a noisy input $\boldsymbol{x}_t$ is generated according to the forward diffusion process. The noise prediction network $\boldsymbol{\epsilon}_{\boldsymbol{\theta}}$ is then trained to minimize the mean squared error (MSE) between its prediction and the true noise:
\begin{equation}
\label{eq:noise_loss}
\mathcal{L}_{\text{noise}}(\boldsymbol{\theta}) = \mathbb{E}_{\boldsymbol{x}_0, \boldsymbol{\epsilon}, t} \Big[ \big\| \boldsymbol{\epsilon} - \boldsymbol{\epsilon}_{\boldsymbol{\theta}}(\boldsymbol{x}_t, t) \big\|_2^2 \Big],
\end{equation}
where $\boldsymbol{\epsilon} \sim \mathcal{N}(\mathbf{0}, \mathbf{I})$ represents the Gaussian noise added at timestep $t$.  
This objective directly aligns with denoising score matching, providing a concrete and tractable learning target for the diffusion model tailored to SAR images.

The complete training procedure for the SAR diffusion noise prediction network is summarized in Algorithm~\ref{alg:training_SAR_simple}.
\begin{algorithm}[t]
\caption{Training Procedure for SAR Diffusion Noise Prediction Network}\label{alg:training_SAR_simple}
\begin{algorithmic}[1]
\REQUIRE A complex-valued SAR image $\boldsymbol{x}_0$ (converted to two-channel real-valued representation), Architecture of the noise prediction network $\boldsymbol{\epsilon}_{\boldsymbol{\theta}}$, number of diffusion timesteps $T$, learning rate $\eta$, noise schedule $\{\alpha_t\}_{t=1}^T$
\STATE Initialize network parameters $\boldsymbol{\theta}$
    \REPEAT
    \STATE Sample a timestep $t \sim \mathcal{U}\{1,\cdots, T\}$
    \STATE Sample Gaussian noise $\boldsymbol{\epsilon} \sim \mathcal{N}(\mathbf{0}, \mathbf{I})$
    \STATE Generate noisy SAR image $\boldsymbol{x}_t = \sqrt{\alpha_t} \boldsymbol{x}_0 + \sqrt{1-\alpha_t} \boldsymbol{\epsilon}$
    \STATE Predict noise $\hat{\boldsymbol{\epsilon}} = \boldsymbol{\epsilon}_{\boldsymbol{\theta}}(\boldsymbol{x}_t, t)$
    \STATE Compute loss $\mathcal{L}_{\text{noise}} \gets$ \eqref{eq:noise_loss}
    \STATE Update network parameters $\boldsymbol{\theta} \gets \boldsymbol{\theta} - \eta \nabla_{\boldsymbol{\theta}} \mathcal{L}_{\text{noise}}$
    \UNTIL converged
\ENSURE Trained  noise prediction network $\boldsymbol{\epsilon}_{\boldsymbol{\theta}}$
\end{algorithmic}
\end{algorithm}

\subsection{Prior Utilization in SAR Imaging}
During reconstruction, the posterior sampling integrates both the likelihood term, derived from the SAR measurement model, and the learned diffusion prior.  
According to the SAR signal model in \eqref{vector_model}, the log-likelihood term is defined as
\[
\mathcal{L}(\boldsymbol{z}_{k}^{(t)}; \boldsymbol{y}) = \left\Vert \boldsymbol{y} - \boldsymbol{A}\boldsymbol{z}_{k}^{(t)} \right\Vert_{2}^{2}.
\]
Using Wirtinger calculus for complex-valued variables, the gradient of $\mathcal{L}$ with respect to the conjugate variable $\boldsymbol{z}^{\ast}$ is
\begin{equation}
\label{eq:Wirtinger}
\nabla_{\boldsymbol{z}^{\ast}} \mathcal{L}\big(\boldsymbol{z}_{k}^{(t)}; \boldsymbol{y}\big)
= \boldsymbol{A}^{\mathrm{H}}\big(\boldsymbol{A}\boldsymbol{z}_{k}^{(t)} - \boldsymbol{y}\big),
\end{equation}
which guides the likelihood-based Langevin updates.  

The prior term is incorporated via the pretrained noise prediction network $\boldsymbol{\epsilon}_{\boldsymbol{\theta}}$, which encodes the SAR data manifold and provides denoising directions for each intermediate state in the diffusion process.  
By alternating between likelihood-driven Langevin steps and prior-guided diffusion updates, the algorithm efficiently explores the posterior distribution while enforcing both data fidelity and SAR-specific priors.

The complete asymptotic sampling procedure is summarized in Algorithm~\ref{alg:Alternating Sampling}. A stochastic prior sampling version of SGS-DDPM~\cite{xu2024provably, 10541919} can likewise be adopted.
\begin{algorithm}[t]
\caption{Split Gibbs Sampler with DDIM (SGS-DDIM) for SAR Imaging}\label{alg:Alternating Sampling}
\begin{algorithmic}[1]
\REQUIRE SAR echo data $\boldsymbol{y}$, measurement model $\mathcal{L}$, pretrained noise predictor $\boldsymbol{\epsilon}_{\boldsymbol{\theta}}$, annealing schedule $\lambda_{k}$, Langevin stepsize $\kappa$
\STATE $\boldsymbol{x}_{0} \sim \mathcal{N} \left(\boldsymbol{0}, \frac{\lambda_{0}}{4} \boldsymbol{I} \right)$
\FOR{$k = 1, \dots, K$} 
    \STATE Set $\boldsymbol{z}_{k}^{(0)} = \boldsymbol{x}_{k-1}$
    \FOR{$t = 1, \dots, T$} 
        \STATE Update $\boldsymbol{z}_{k}^{(t)} \gets$ \eqref{Langevin_Dynamics_EI}
    \ENDFOR
    \STATE Set $\boldsymbol{z}_{k} = \boldsymbol{z}_{k}^{(T)}$
    \STATE $\boldsymbol{\zeta}^{(N)} \sim \mathcal{N}(\boldsymbol{0}, \boldsymbol{I})$
    \FOR{$i = N-1, \dots, 0$} 
        \STATE Update $\boldsymbol{\zeta}^{(i)} \gets$ \eqref{deterministic_sample}
    \ENDFOR
    \STATE $\boldsymbol{x}_{k} = \boldsymbol{z}_{k} + \boldsymbol{\zeta}^{(0)}$
\ENDFOR
\ENSURE $\boldsymbol{x}_{K}$
\end{algorithmic}
\end{algorithm}

\section{Experiments}\label{experiments}
\subsection{Baseline and Evaluation Metrics}
To comprehensively evaluate the effectiveness of the asymptotic methods, SGS-DDPM and SGS-DDIM, we compare them with representative baselines covering different categories of reconstruction strategies:  
\begin{enumerate}
    \item \textbf{MF}: a matched filter reconstruction without any prior information, serving as a non-regularized baseline;  
    \item \textbf{FISTA}~\cite{beck2009fast} and \textbf{ADMM}~\cite{boyd2011distributed}: optimization-based approaches that incorporate handcrafted sparsity priors to promote sparse representations of the scene;  
    \item \textbf{DPS}~\cite{chung2022diffusion}: a posterior sampling method that leverages diffusion models as learned priors and implements posterior sampling through a guidance method.  
\end{enumerate}
These baselines represent reconstructions with no priors, with analytical priors, and with diffusion-based priors, respectively, thus enabling a fair and comprehensive comparison and demonstrating the necessity of utilizing priors.
 
We use six metrics to comprehensively evaluate the above methods, which assess both reconstruction fidelity, structural preservation ability, and perceived similarity.
Four widely adopted image reconstruction metrics—Normalized Mean Square Error (NMSE), Peak Signal-to-Noise Ratio (PSNR), Structural Similarity Index (SSIM) \cite{li2024sar}, and Learned Perceptual Image Patch Similarity (LPIPS)\cite{zhang2018unreasonable}—are employed to evaluate pixel-level accuracy and perceptual consistency.  
In addition, for numerical simulations, we further assess the sidelobe characteristics using the Modified Peak Sidelobe Ratio (MPSLR) and Modified Integrated Sidelobe Ratio (MISLR), which are adapted for distributed targets.

\textbf{NMSE} quantifies the normalized reconstruction error relative to the true signal energy:
\begin{equation}
\label{eq:NMSE}
    \text{NMSE} = 
    \frac{\left\Vert \boldsymbol{X} - \hat{\boldsymbol{X}} \right\Vert_{F}^{2}}
    {\left\Vert \boldsymbol{X} \right\Vert_{F}^{2}},
\end{equation}
where $\boldsymbol{X}$ and $\hat{\boldsymbol{X}}$ denote the ground-truth and reconstructed backscattering matrices, respectively, and $\Vert \cdot \Vert_{F}$ is the Frobenius norm.

\textbf{PSNR} measures signal fidelity on a logarithmic scale, emphasizing deviations in high-intensity regions:
\begin{equation}
\label{eq:PSNR}
    \text{PSNR} = 
    20 \log_{10} \left( 
    \frac{\max \left( \boldsymbol{I}_{\boldsymbol{X}} \right)}
    {\text{RMSE}}
    \right),
\end{equation}
where $\boldsymbol{I}_{\boldsymbol{X}} = \vert \boldsymbol{X}\vert$, and the root mean square error (RMSE) is defined as
$\text{RMSE} = 
\sqrt{\tfrac{1}{PQ}
\left\Vert 
\boldsymbol{I}_{\boldsymbol{X}} - 
\boldsymbol{I}_{\hat{\boldsymbol{X}}}
\right\Vert_{F}^{2}}$.

\textbf{SSIM} evaluates perceptual similarity by jointly accounting for luminance, contrast, and structural information:
\begin{equation}
\label{eq:SSIM}
\text{SSIM} =
\frac{
\left( 2\mu_{\boldsymbol{I}_{\boldsymbol{X}}}
\mu_{\boldsymbol{I}_{\hat{\boldsymbol{X}}}} + C_{1} \right)
\left( 2\sigma_{\boldsymbol{I}_{\boldsymbol{X}}
\boldsymbol{I}_{\hat{\boldsymbol{X}}}} + C_{2} \right)}
{\left( \mu_{\boldsymbol{I}_{\boldsymbol{X}}}^{2} +
\mu_{\boldsymbol{I}_{\hat{\boldsymbol{X}}}}^{2} + C_{1} \right)
\left( \sigma_{\boldsymbol{I}_{\boldsymbol{X}}}^{2} +
\sigma_{\boldsymbol{I}_{\hat{\boldsymbol{X}}}}^{2} + C_{2} \right)},
\end{equation}
where $\mu$ and $\sigma^2$ denote the mean and variance of image intensity, respectively, and $\sigma_{\boldsymbol{I}_{\boldsymbol{X}} \boldsymbol{I}_{\hat{\boldsymbol{X}}}}$ is the cross-covariance term.

\textbf{LPIPS} assesses perceptual similarity based on feature representations extracted from pretrained deep networks, offering a metric more aligned with human visual perception:
\begin{equation}
\label{eq:LPIPS}
    \text{LPIPS} = 
    \sum_{l} \frac{1}{PQ} 
    \sum_{p,q}
    \left\Vert 
    w_{l} \odot 
    \big( 
    \phi_{l}(\boldsymbol{I}_{\boldsymbol{X}})_{pq} - 
    \phi_{l}(\boldsymbol{I}_{\hat{\boldsymbol{X}}})_{pq}
    \big)
    \right\Vert_{2}^{2},
\end{equation}
where $\phi_{l}(\cdot)$ denotes the feature map from the $l$-th layer of a pretrained backbone (e.g., AlexNet or VGG), $w_{l}$ is a learned channel-wise weight, and $\odot$ denotes element-wise multiplication.

\vspace{0.3em}
\noindent
\textbf{Sidelobe Metrics.}  
Since our scenes contain distributed rather than isolated point targets, the conventional Peak Sidelobe Ratio (PSLR) and Integrated Sidelobe Ratio (ISLR) are not directly applicable.  
However, in numerical simulations, the target support set $\Lambda$ is known, allowing us to define modified versions that measure the residual energy and peaks outside the true support.

\textbf{MPSLR} quantifies the relative amplitude between the main-lobe and sidelobe regions:
\begin{equation}
\label{eq:MPSLR}
    \text{MPSLR} = 
    20 \log_{10}
    \left(
    \frac{A_{\text{main}}}{A_{\text{side}}}
    \right),
\end{equation}
where 
$A_{\text{main}} = 
\max\limits_{(i,j) \in \Lambda}
\left \vert \hat{\boldsymbol{X}}_{ij} \right \vert$
is the main-lobe peak within the support region, and
$A_{\text{side}} =
\max\limits_{(i,j) \in \Lambda^{c}}
\left \vert \hat{\boldsymbol{X}}_{ij} \right \vert$
is the maximum sidelobe amplitude outside the support.

\textbf{MISLR} evaluates the energy ratio between the main-lobe and sidelobe regions:
\begin{equation}
\label{eq:MISLR}
    \text{MISLR} = 
    10 \log_{10}
    \left(
    \frac{E_{\text{main}}}{E_{\text{side}}}
    \right),
\end{equation}
where 
$E_{\text{main}} =
\sum\limits_{(i,j) \in \Lambda}
\left \vert \hat{\boldsymbol{X}}_{ij} \right\vert^{2}$
is the total main-lobe energy, and
$E_{\text{side}} =
\sum\limits_{(i,j) \in \Lambda^{c}}
\left\vert \hat{\boldsymbol{X}}_{ij} \right\vert^{2}$
is the total sidelobe energy.

\vspace{0.3em}
\noindent
For each evaluation metric, all competing methods are independently repeated 20~times under the same experimental conditions, and the averaged results are reported for quantitative comparison.

\subsection{Datasets and Experimental Setup}
\textbf{Datasets.}
We evaluate all the methods on two datasets: a simulated dataset and a real SAR dataset.  
The simulated dataset is constructed by assigning a random phase to each pixel of the MNIST images \cite{lecun2002gradient}, formulated as
\begin{equation}
\label{mnist}
    \boldsymbol{X} =  \boldsymbol{I}_{0} \odot e^{j \boldsymbol{\phi} },
\end{equation}
where $\boldsymbol{X} \in \mathbb{C}^{32 \times 32}$, $\boldsymbol{I}_{0} \in \mathbb{R}^{32 \times 32}$ represents the normalized intensity image, $\boldsymbol{\phi}$ is a random phase matrix with independent elements drawn from a uniform distribution $\phi_{pq} \sim \mathcal{U}[-\pi, \pi)$, and $\odot$ denotes the Hadamard product.  

The real dataset consists of single-look complex (SLC) data acquired by the Sentinel-1A SAR sensor. The raw data are divided into multiple complex patches of size $256 \times 256$. The radar system parameters are summarized in Table~\ref{radar_parameters}.  

\begin{table}[ht]
\centering
\caption{Radar parameters}
\begin{tabular}{cccccc}
\toprule
Parameter          & Unit  & Value    & Parameter           & Unit & Value \\
\midrule
Orbital altitude   &  km   &  693     &  Orbital velocity   & km/s & 7.12 \\
Carrier frequency  &  GHz  & 5.405    &  Sampling rate      & MHz  & 64.35 \\
PRF                &  kHz  & 1.717    &  Bandwidth          & MHz  & 56.5 \\
Range resolution   &   m   &   5      & Azimuth resolution  &  m   & 20 \\
\bottomrule
\end{tabular}
\label{radar_parameters}
\end{table}

\textbf{Hyperparameters Setup.}
Following Algorithm~\ref{alg:Alternating Sampling}, we employ the following annealing schedule for $\lambda_{k}$:
\[
\kappa = 0.8 \cdot \min \left\{ \lambda_{k}^{2},  0.15 \right\}, \quad
\lambda_{k} = \lambda_{0} \left( \frac{\lambda_{K}}{\lambda_{0}} \right)^{\max \left(0, \frac{k-K_{0}}{K-K_{0}} \right)}.
\]
For the MNIST dataset, we set $\lambda_{0} = 0.35$ and $\lambda_{K} = 0.05$, while for Sentinel-1A, we use $\lambda_{0} = 0.45$ and $\lambda_{K} = 0.15$. For both datasets, the remaining hyperparameters are fixed as $K_{0} = 15$, and $K = 60$.  

\textbf{Network Architecture.}
For the MNIST dataset, the network comprises two encoders with depths $\{5,4\}$, a middle layer of depth~4, and two decoders with depths $\{4,4\}$.
For the Sentinel-1A dataset, the architecture includes three encoders with depths $\{1,2,4\}$, a middle layer of depth~6, and three decoders with depths $\{4,2,1\}$.

All training experiments are conducted on two NVIDIA RTX~A6000 GPUs.

\subsection{Experiment Results}
\begin{figure*}[!t]
    \centering
    \subfloat[Ground Truth]{
    \includegraphics[width=0.95in]{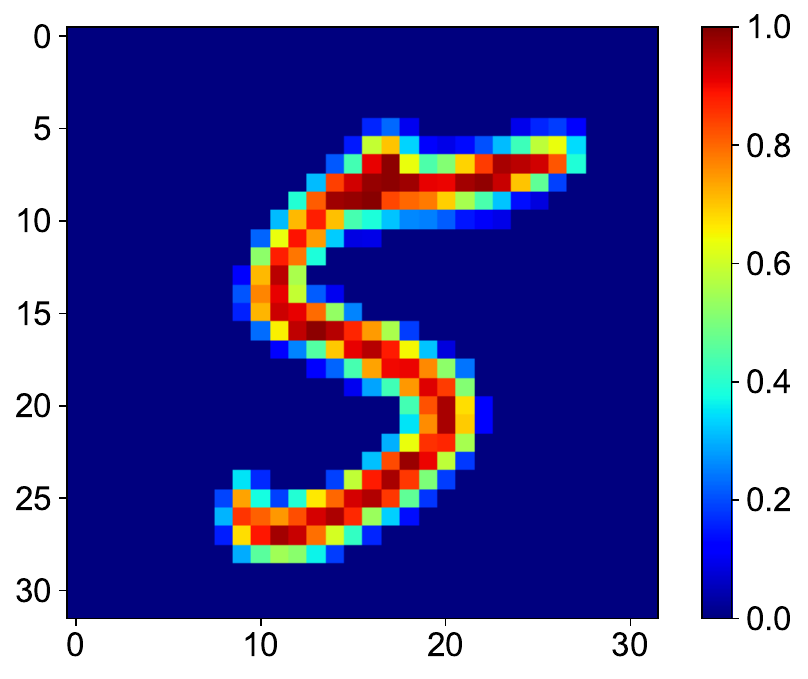}}\hspace{2pt}
    \subfloat[MF]{\includegraphics[width=0.95in]{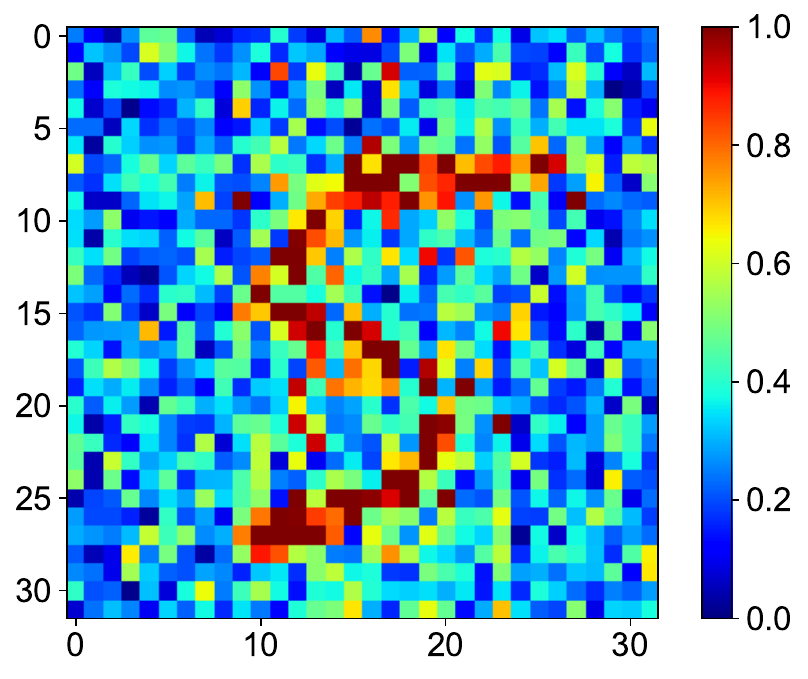}}\hspace{2pt}
    \subfloat[FISTA]{\includegraphics[width=0.95in]{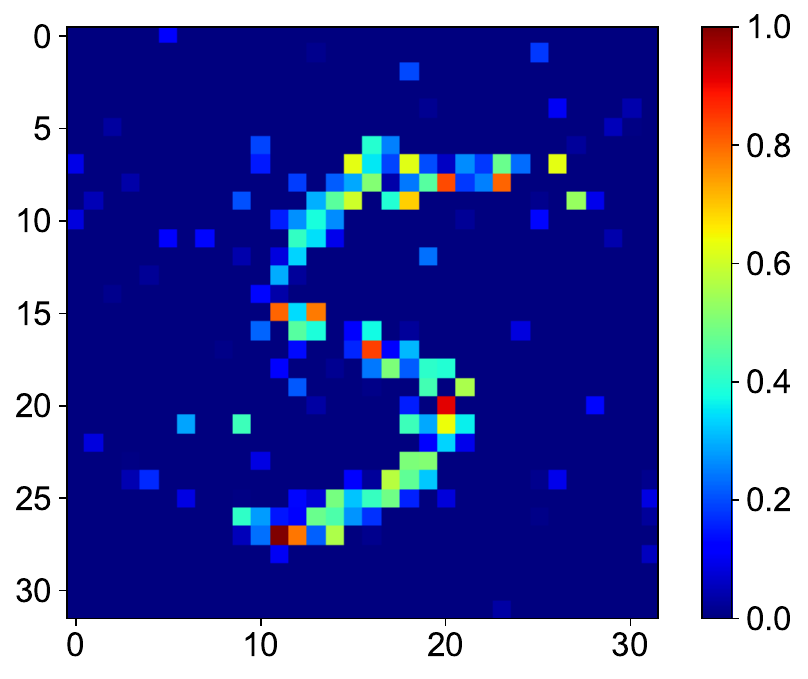}}\hspace{2pt}
    \subfloat[ADMM]{\includegraphics[width=0.95in]{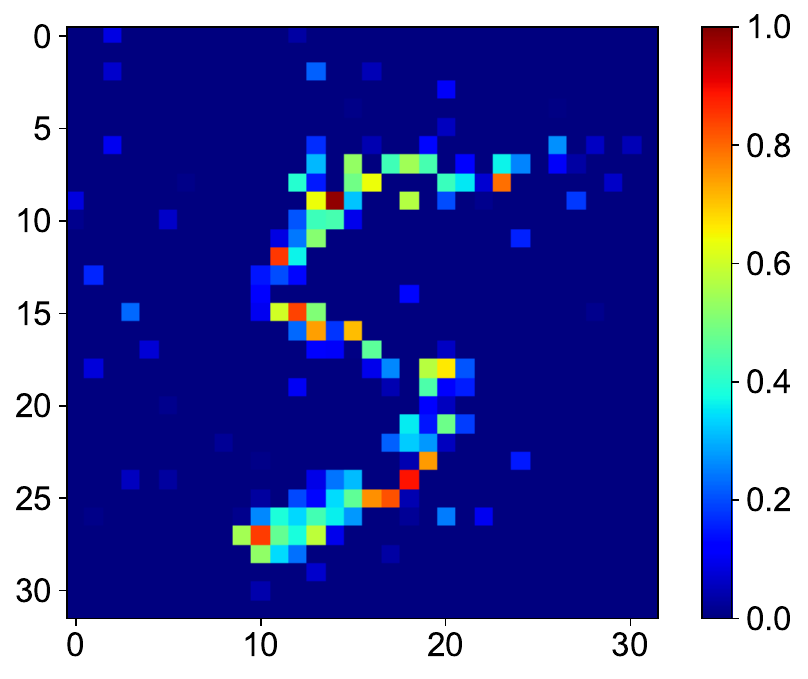}}\hspace{2pt}
    \subfloat[DPS]{\includegraphics[width=0.95in]{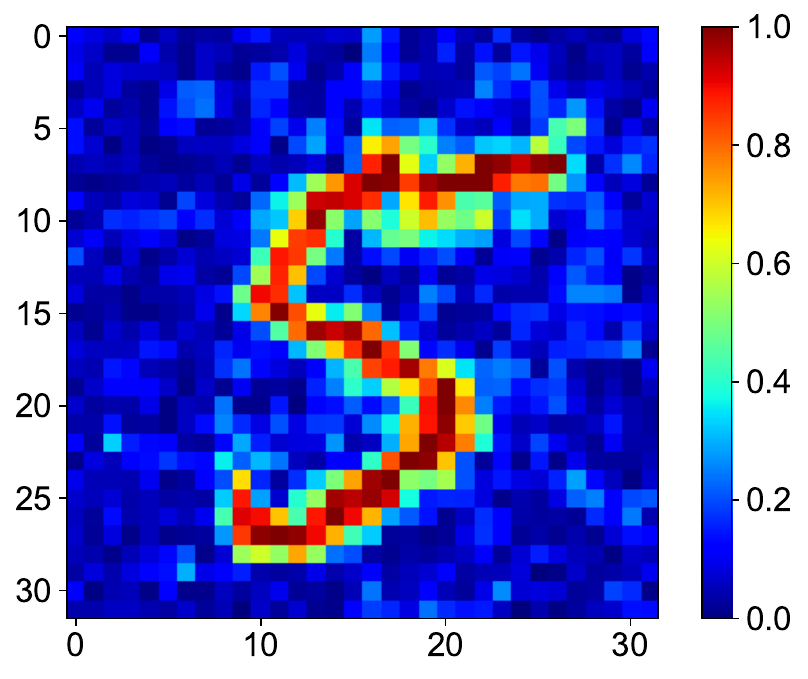}}\hspace{2pt}
    \subfloat[SGS-DDPM]{\includegraphics[width=0.95in]{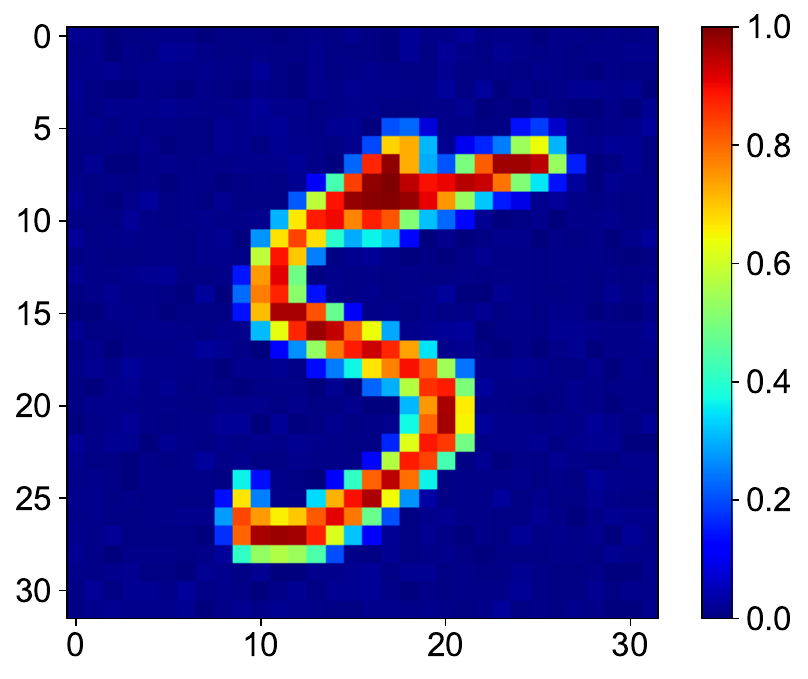}}\hspace{2pt}
    \subfloat[SGS-DDIM]{\includegraphics[width=0.95in]{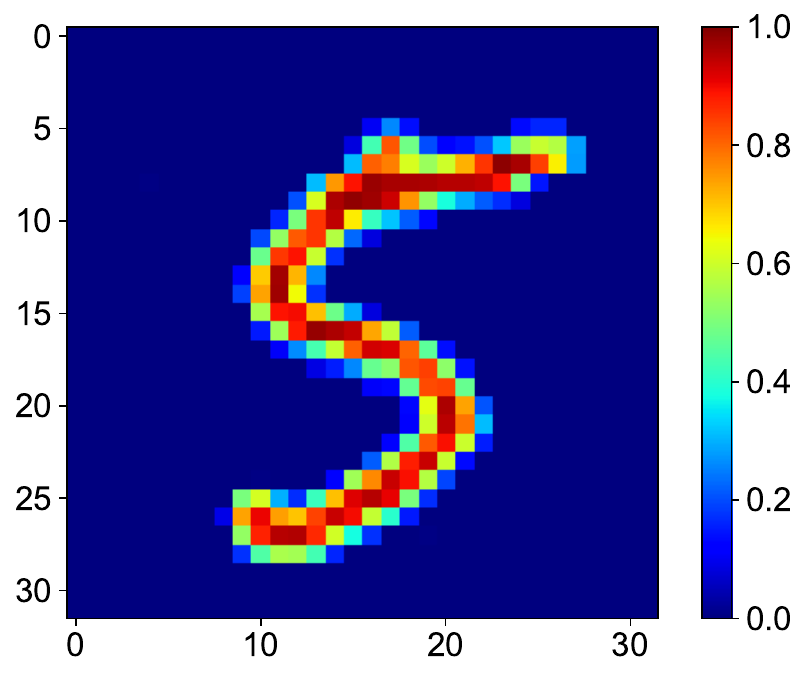}}
    \caption{Reconstruction results comparison under 2 dB SNR and 24 sampling points. Single recovery results of each algorithm.}
    \label{fig:simu_recon_results}
\end{figure*}

\begin{table*}[!t]
\centering
\caption{Quantitative evaluation under 2 dB SNR and 24 sampling points. NMSE, PSNR, MPLSR, and MILSR are reported in dB. \textbf{Bold}:best; \underline{Underline}: second best.}
\begin{tabular}{ccccccc}
\toprule
Method & NMSE $\downarrow$ & PSNR $\uparrow$ & SSIM $\uparrow$ & LPIPS $\downarrow$ & MPSLR $\downarrow$ & MISLR $\downarrow$\\
\midrule
MF      & 2.71     & 8.83    & 0.299    & 0.689    & -4.46    & -0.189 \\
FISTA   & -2.28    & 13.4    & 0.347    & 0.324    & -7.73    & -12.1   \\
ADMM    & -2.23    & 13.4    & 0.318    & 0.348    & -8.04    & -12.2   \\
DPS     & -3.02    & 15.7    & 0.511    & 0.435    & -5.37    & -6.68   \\
SGS-DDPM& \textbf{-5.49}   & \textbf{23.5}   & \underline{0.789}   & \underline{0.0360}   & \underline{-11.0}  & \underline{-23.7}  \\
SGS-DDIM& \underline{-5.31}   & \underline{23.2}   & \textbf{0.900}   & \textbf{0.0321}  & \textbf{-11.2}  & \textbf{-26.0}  \\
\bottomrule
\end{tabular}
\label{tab:simu_quan_eval_snr_sr}
\end{table*}
We first validate all the methods on simulated complex-valued data under a fixed signal-to-noise ratio (SNR) of 2~dB and a sampling rate of 24~points in both dimensions.  
Fig.~\ref{fig:simu_recon_results} presents the qualitative reconstruction results from a single realization.  
The MF result is severely degraded by strong sidelobes that obscure the target structure.  
FISTA and ADMM, benefiting from handcrafted sparsity priors, effectively suppress sidelobes but suffer from partial target loss and blurred edges.  
The DPS method yields improved reconstruction owing to the learned diffusion prior; however, noticeable sidelobes and smeared boundaries remain.  
In contrast, asymptotic methods (SGS-DDPM and SGS-DDIM) recover the target region with higher fidelity and finer structural detail, while maintaining strong sidelobe suppression and edge preservation.

Table~\ref{tab:simu_quan_eval_snr_sr} quantitatively summarizes the performance of different methods.  
The SGS-based samplers achieve the lowest NMSE and highest PSNR, indicating their superior numerical fidelity.  
Both SGS-DDPM and SGS-DDIM yield significant improvements of over 7 dB in PSNR compared with the DPS baseline, demonstrating that the split Gibbs sampling framework effectively enhances the utilization of the diffusion prior.  
In terms of perceptual quality, SGS-DDPM and SGS-DDIM substantially reduce the LPIPS score and boost SSIM, showing that they preserve the structural and textural details of the reconstructed targets.  Furthermore, the sidelobe metrics (MPSLR and MISLR) show that SGS-based methods achieve about a 3 dB improvement MPSLR and at least an 11.5 dB improvement in MISLR compared to classical sparse reconstruction methods (FISTA or ADMM), demonstrating their stronger suppression of spurious artifacts and superior discrimination of true scattering regions.

\begin{figure*}[!t]
\centering
\subfloat[NMSE]{\includegraphics[width=2.2in]{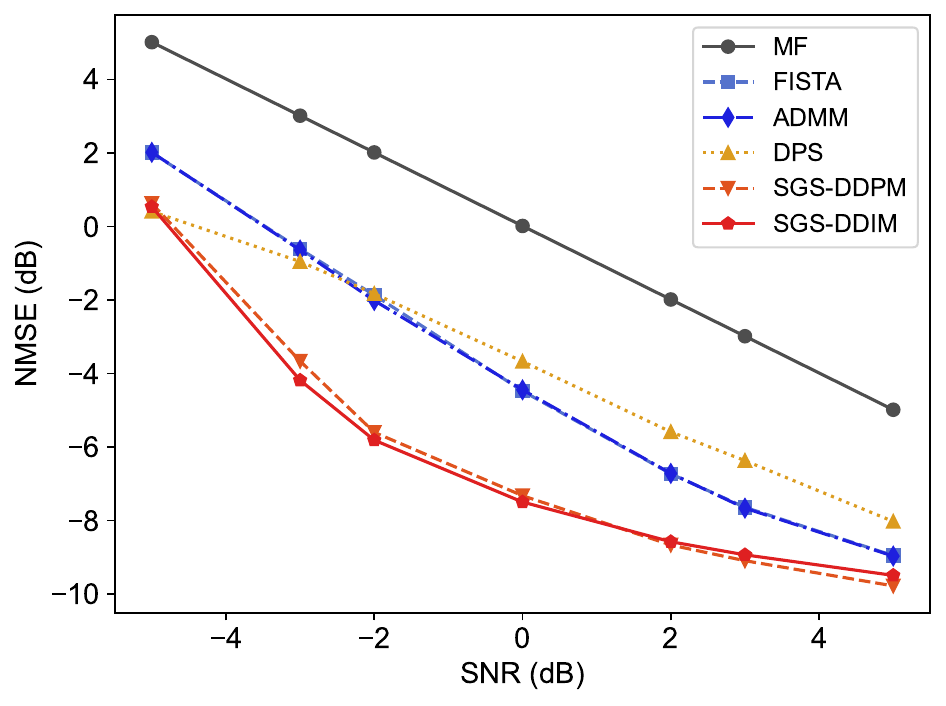}%
\label{nmse_vs_snr}}
\hfil
\subfloat[PSNR]{\includegraphics[width=2.2in]{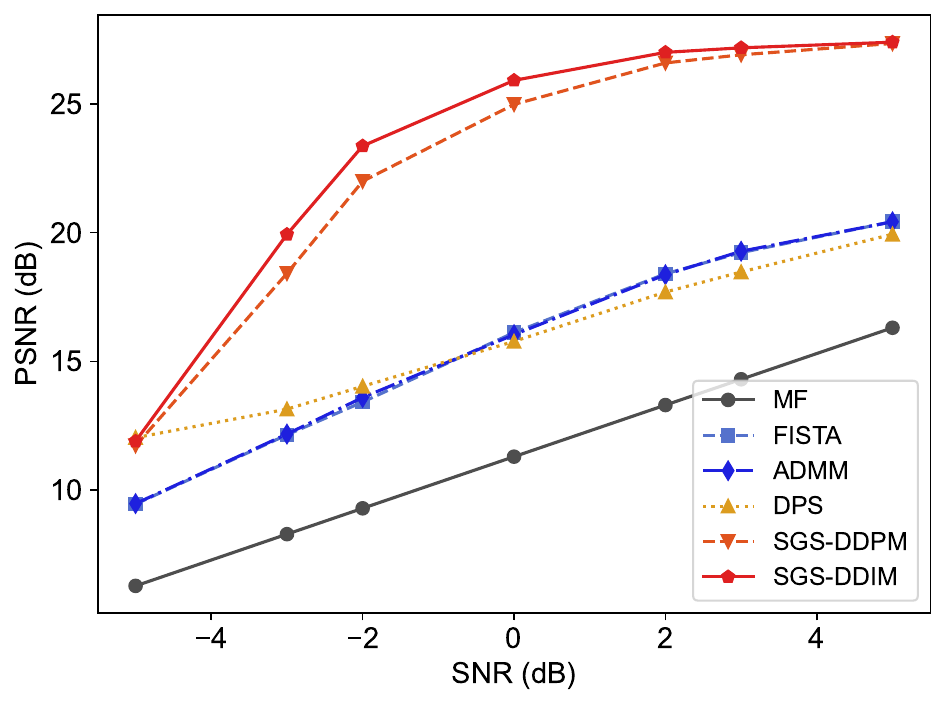}%
\label{psnr_vs_snr}}
\hfil
\subfloat[SSIM]{\includegraphics[width=2.2in]{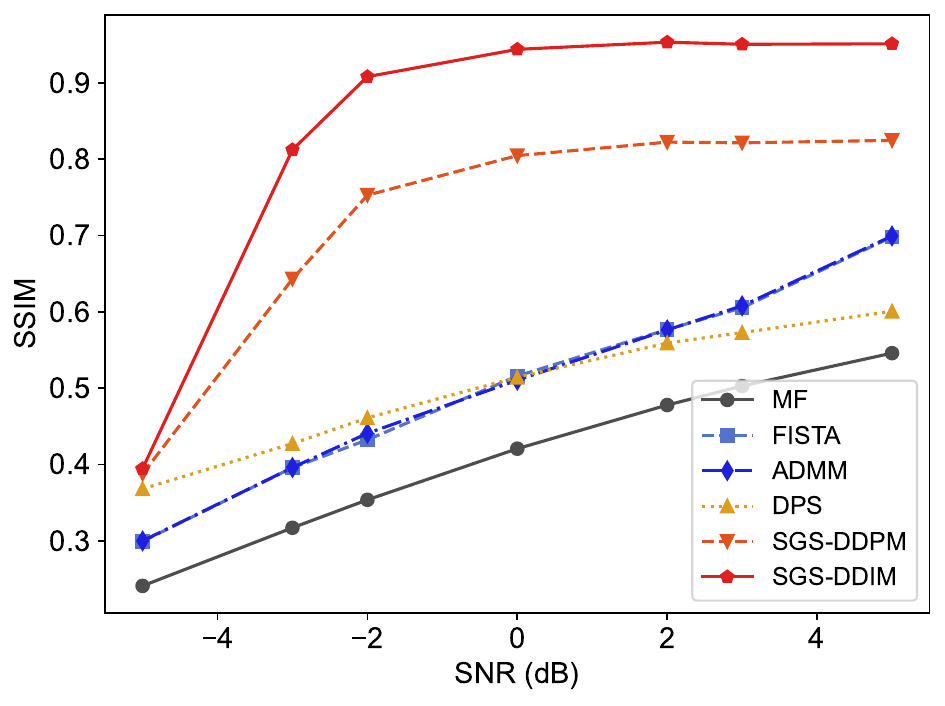}%
\label{lpips_vs_snr}}

\subfloat[LPIPS]{\includegraphics[width=2.2in]{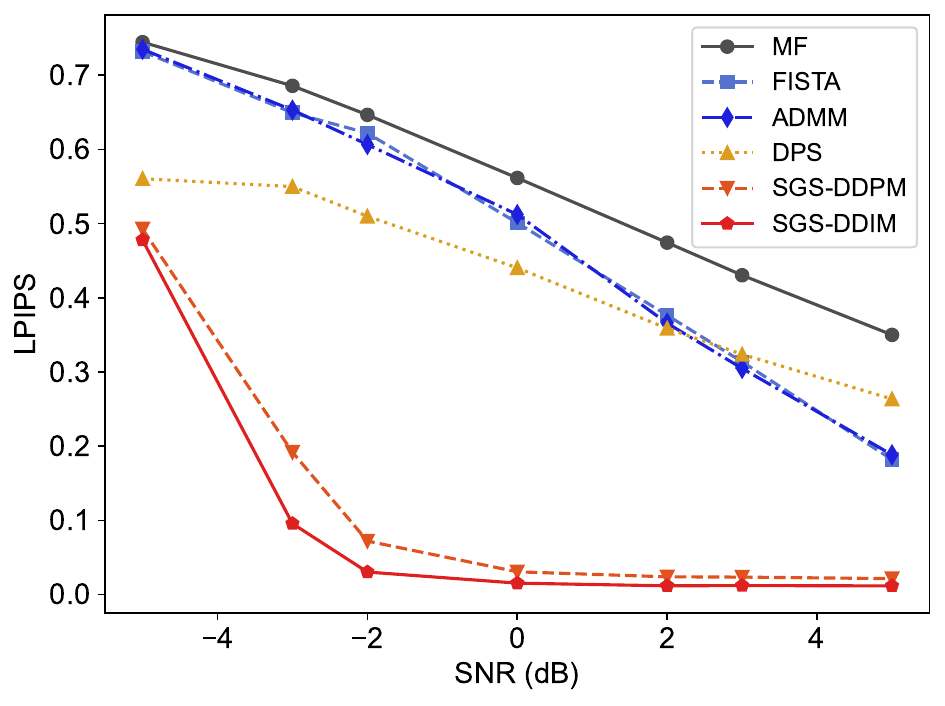}%
\label{ssim_vs_snr}}
\hfil
\subfloat[MPSLR]{\includegraphics[width=2.2in]{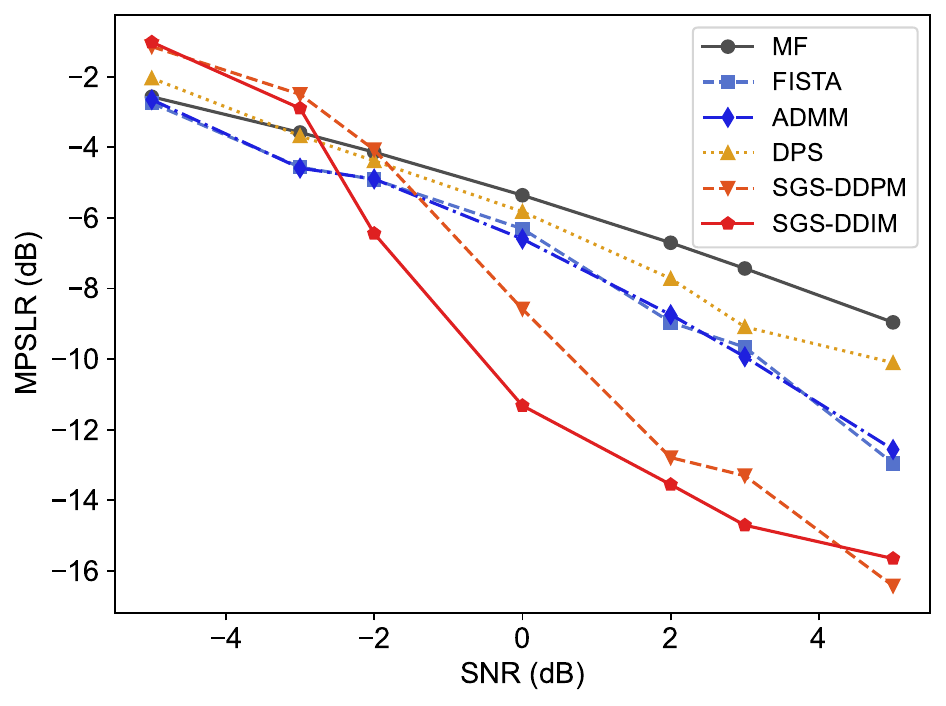}%
\label{mpslr_vs_snr}}
\hfil
\subfloat[MISLR]{\includegraphics[width=2.2in]{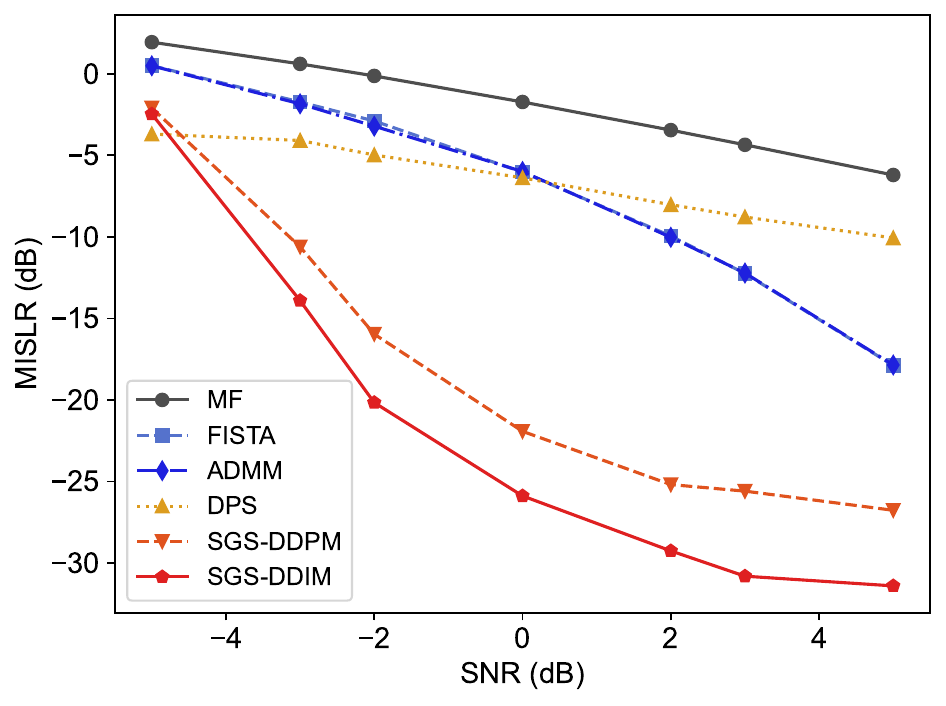}%
\label{mislr_vs_snr}}
\caption{Quantitative reconstruction performance under different SNRs.}
\label{fig:metrics_vs_snr}
\end{figure*}

\begin{figure*}[!t]
\centering
\subfloat[NMSE]{\includegraphics[width=2.2in]{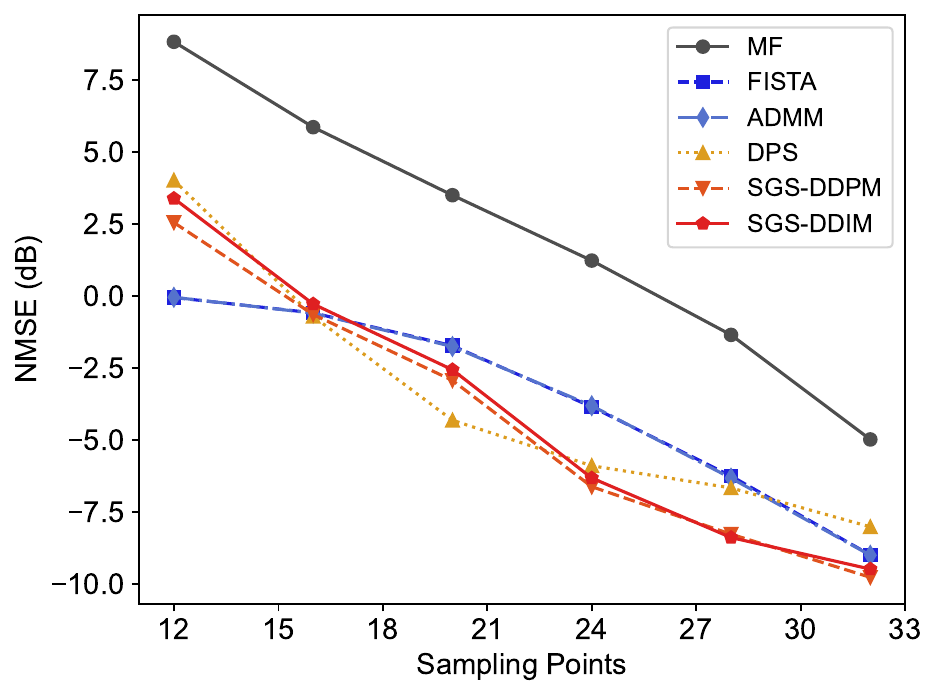}%
\label{nmse_vs_sp}}
\hfil
\subfloat[PSNR]{\includegraphics[width=2.2in]{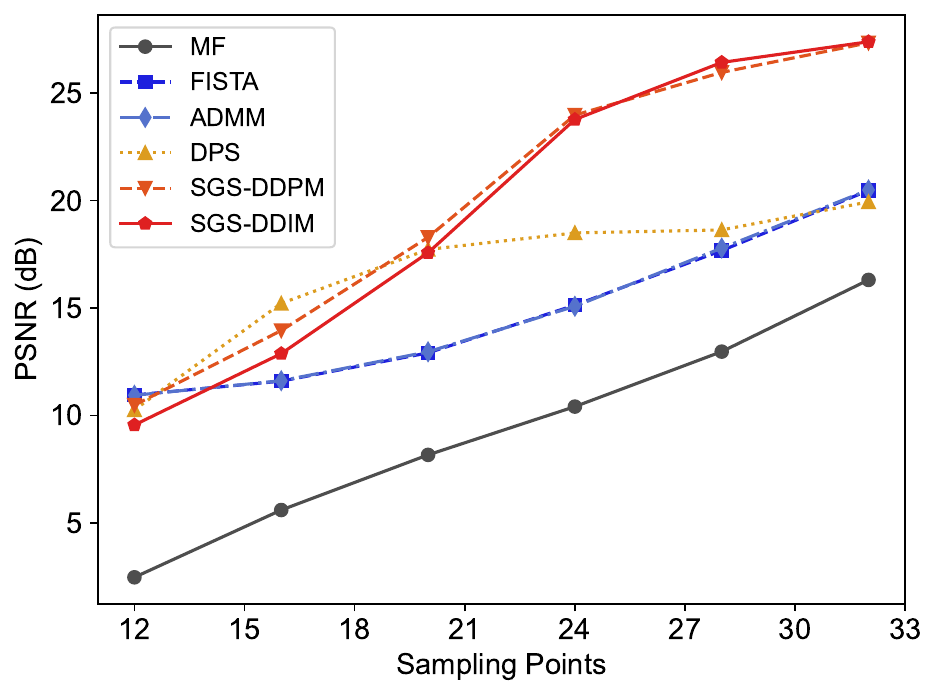}%
\label{psnr_vs_sp}}
\hfil
\subfloat[SSIM]{\includegraphics[width=2.2in]{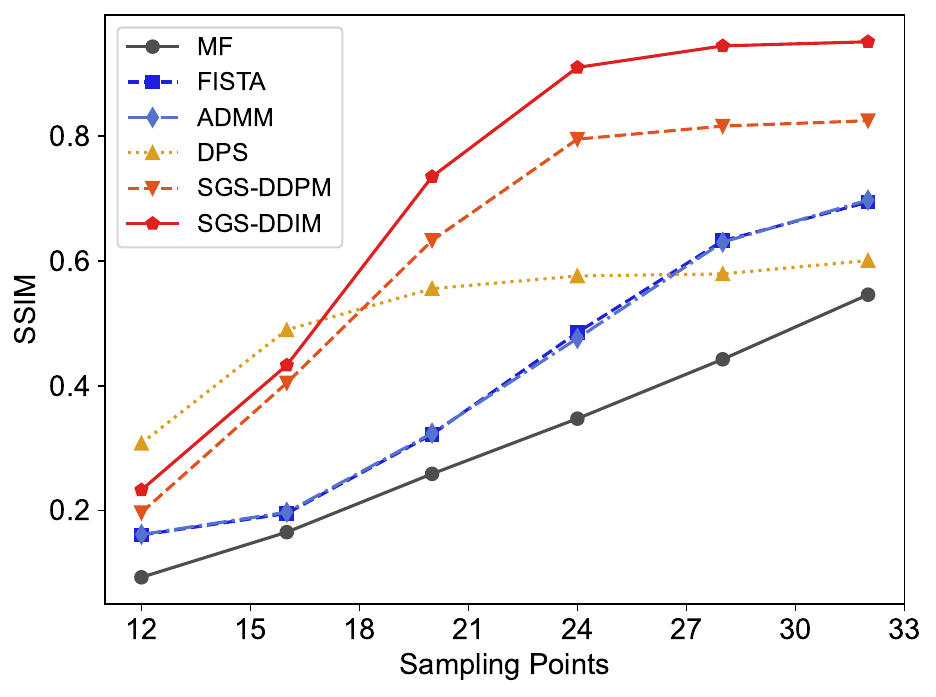}%
\label{ssim_vs_sp}}

\subfloat[LPIPS]{\includegraphics[width=2.2in]{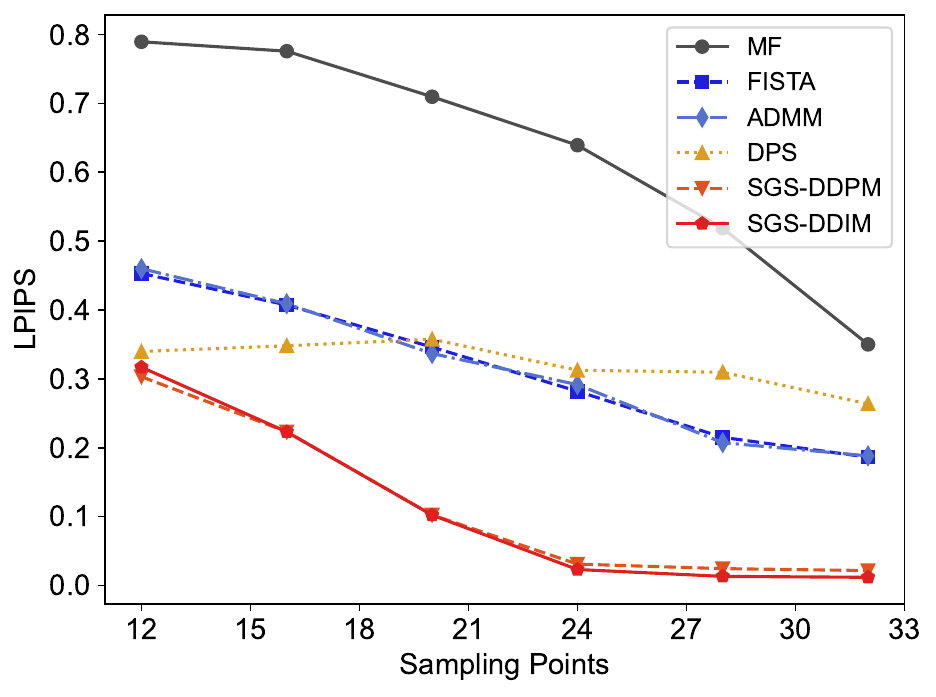}%
\label{lpips_vs_sp}}
\hfil
\subfloat[MPSLR]{\includegraphics[width=2.2in]{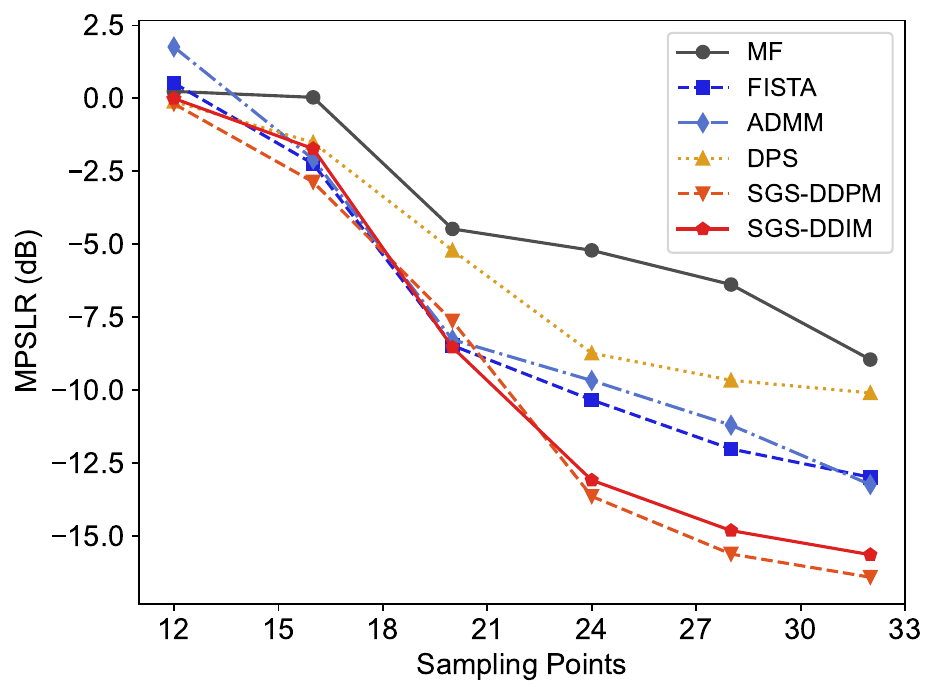}%
\label{mpslr_vs_sp}}
\hfil
\subfloat[MISLR]{\includegraphics[width=2.2in]{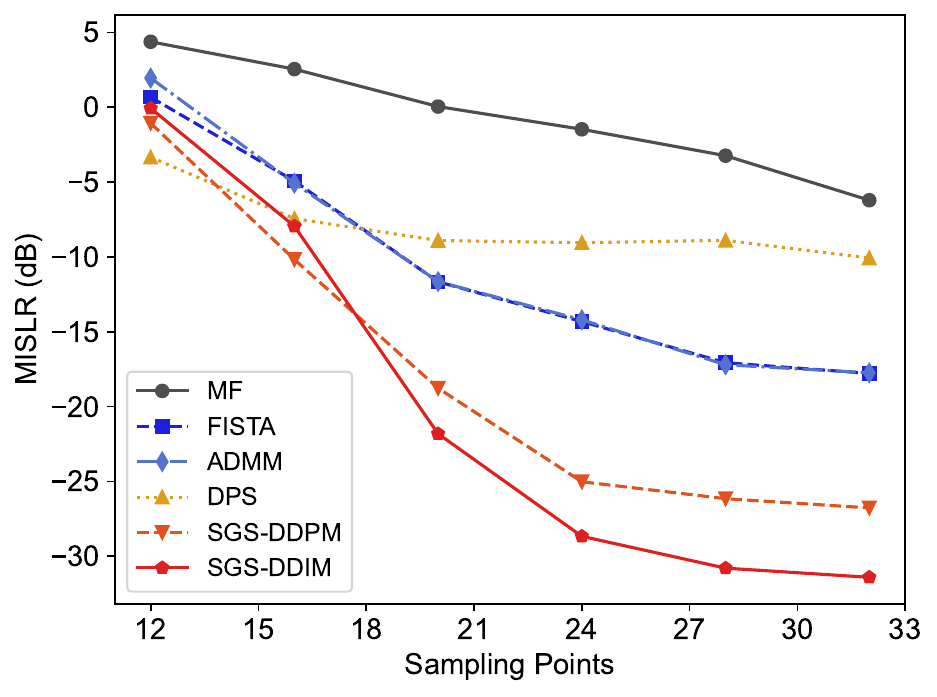}%
\label{mislr_vs_sp}}
\caption{Quantitative reconstruction performance under different sampling points.}
\label{fig:metrics_vs_sp}
\end{figure*}

In the simulation experiments, we further conducted performance evaluations under varying SNRs $\{-5, -3, -2, 0, 2, 3, 5\}$~dB and different sampling rates corresponding to sampling points $\{12, 16, 20, 24, 28, 32\}$. 
For each setting, six performance metrics were plotted as curves versus SNR and sampling points, as shown in Fig~\ref{fig:metrics_vs_snr} and Fig~\ref{fig:metrics_vs_sp}, respectively. 

When $\mathrm{SNR} = -5$~dB or the number of sampling points equals 12, all methods failed to achieve effective reconstruction. 
As the SNR and sampling rate increase, all metrics exhibit consistent improvement. 
Asymptotic maintains superior performance in most cases, 
particularly under low-SNR conditions ($-2$~dB and $-3$~dB) and low sampling rates (16 and 20 points), 
where other methods tend to degrade significantly, while ours still achieves stable and reliable reconstruction.

\begin{figure*}[!t]
    \centering
    \subfloat[Ground Truth]{
    \includegraphics[width=1.7in]{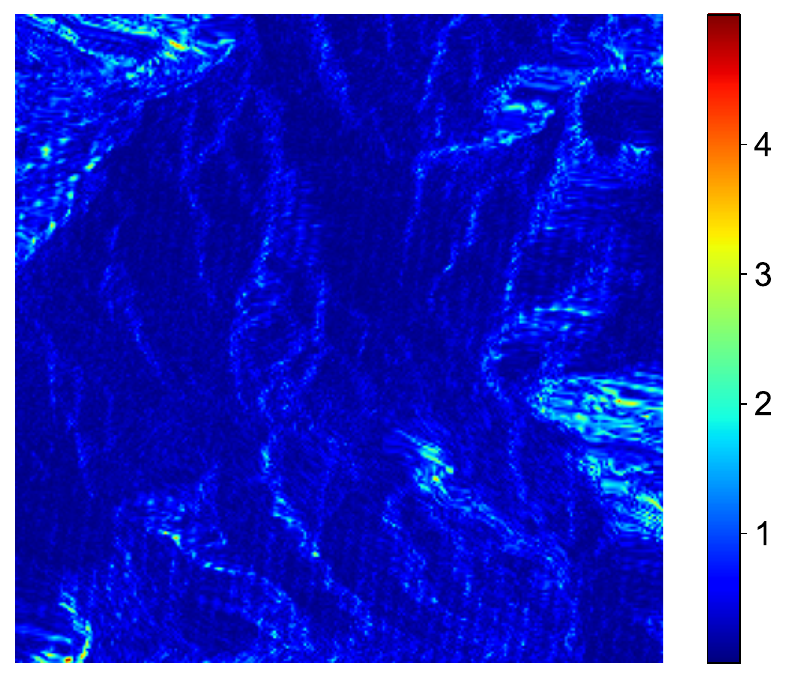}}\hspace{2pt}
    \subfloat[MF]{\includegraphics[width=1.7in]{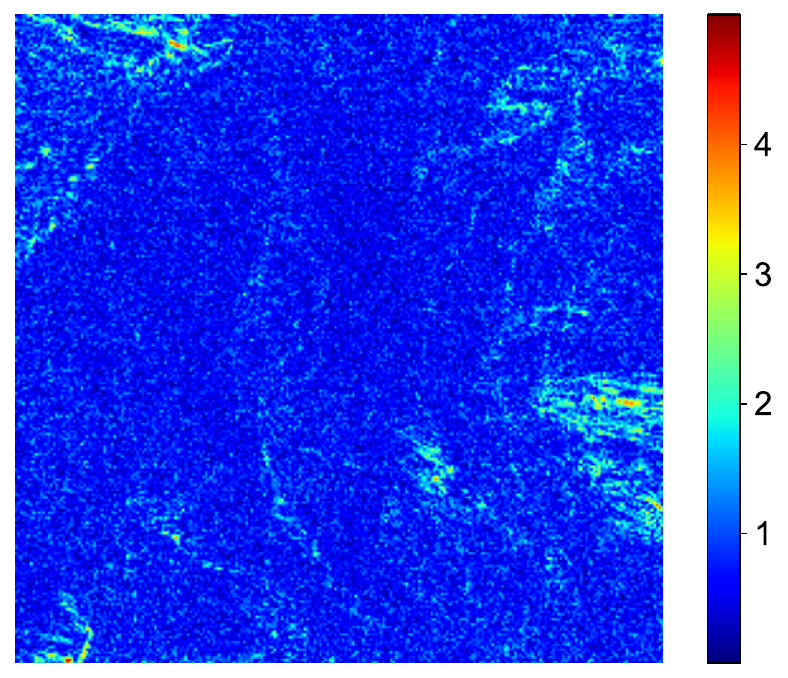}}\hspace{2pt}
    \subfloat[FISTA]{\includegraphics[width=1.7in]{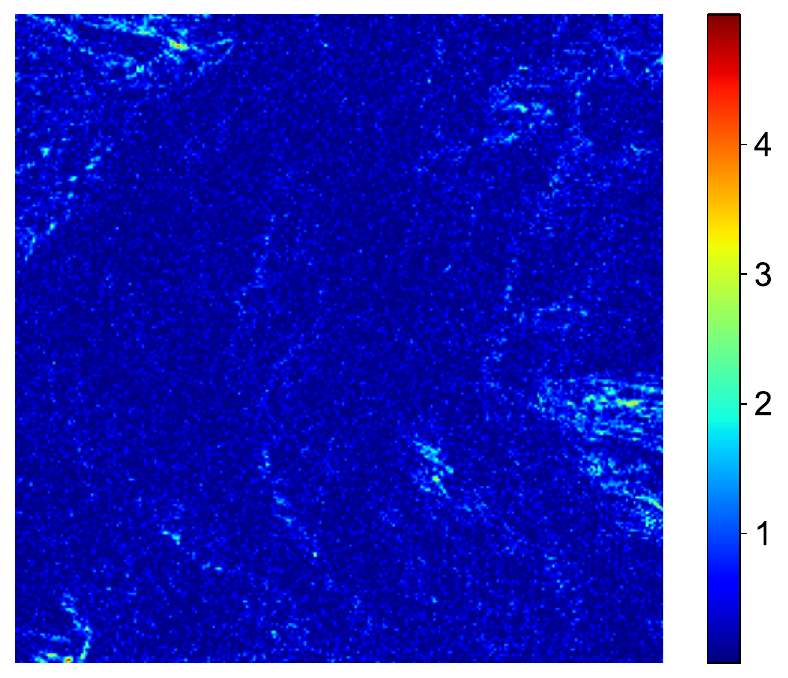}}\hspace{2pt}
    \subfloat[ADMM]{\includegraphics[width=1.7in]{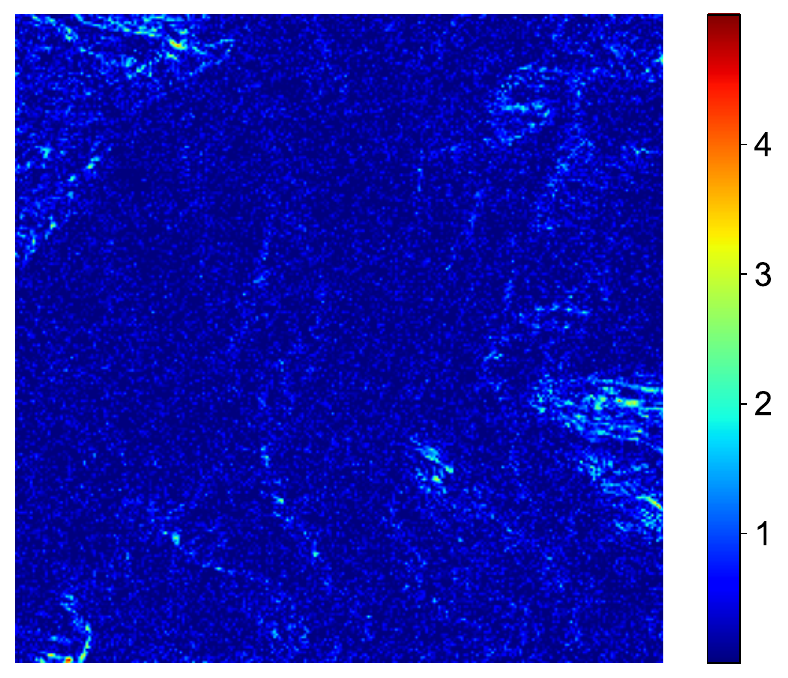}}\hspace{2pt}

    \subfloat[DPS]{\includegraphics[width=1.7in]{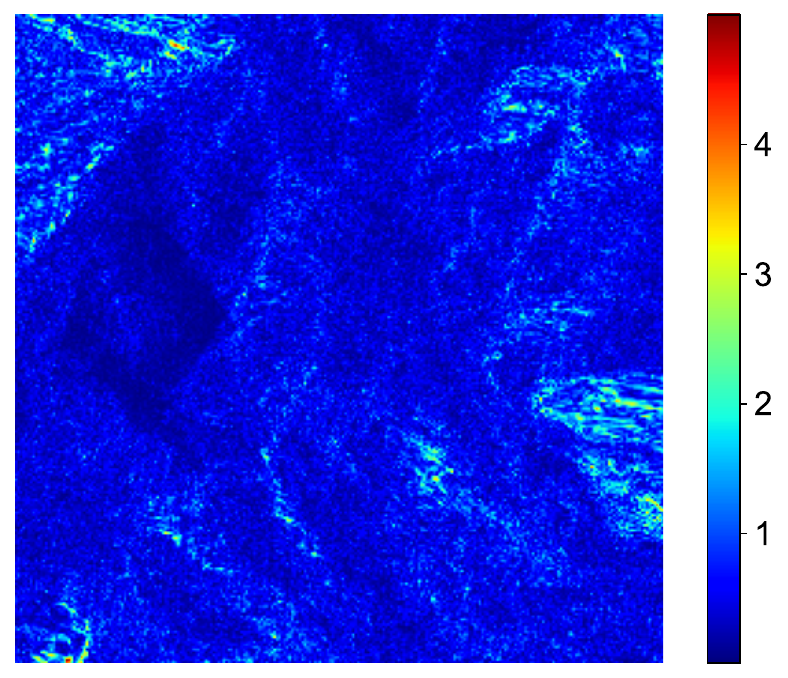}}\hspace{2pt}
    \subfloat[SGS-DDPM]{\includegraphics[width=1.7in]{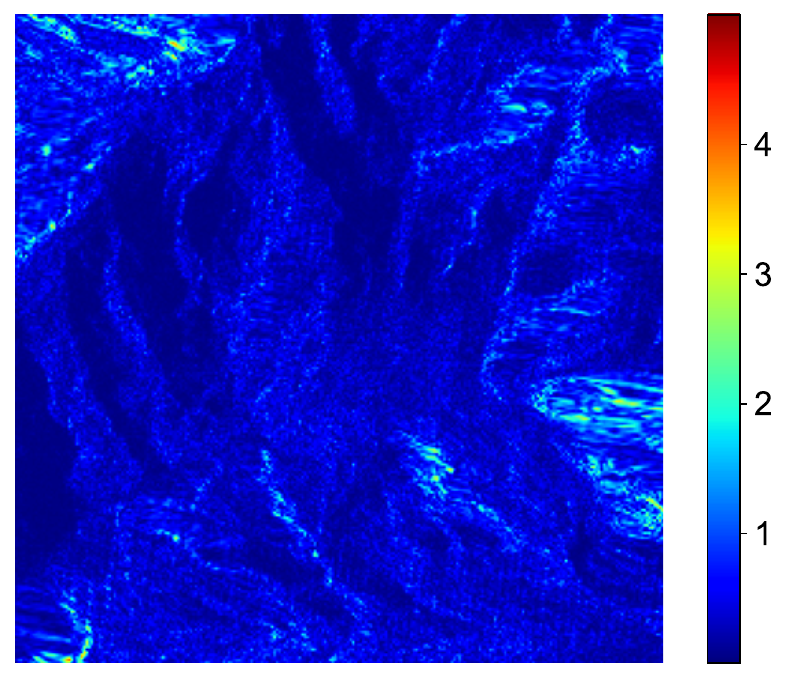}}\hspace{2pt}
    \subfloat[SGS-DDIM]{\includegraphics[width=1.7in]{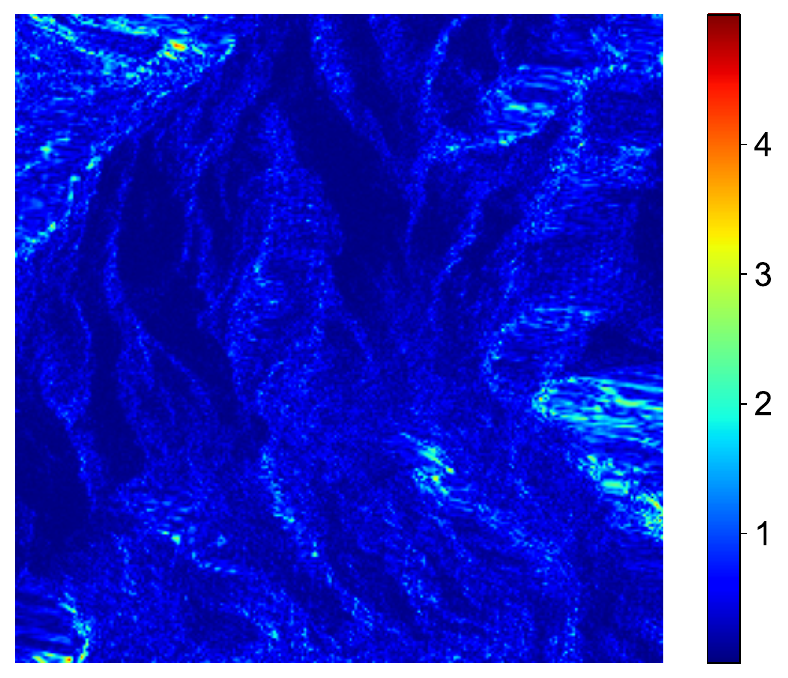}}
    \caption{Reconstruction results comparison under 2 dB SNR and 24 sampling points. Single recovery results of each algorithm.}
    \label{fig:real_scene1_recon_results}
\end{figure*}

\begin{figure*}[!t]
    \centering
    \subfloat[Ground Truth]{
    \includegraphics[width=1.7in]{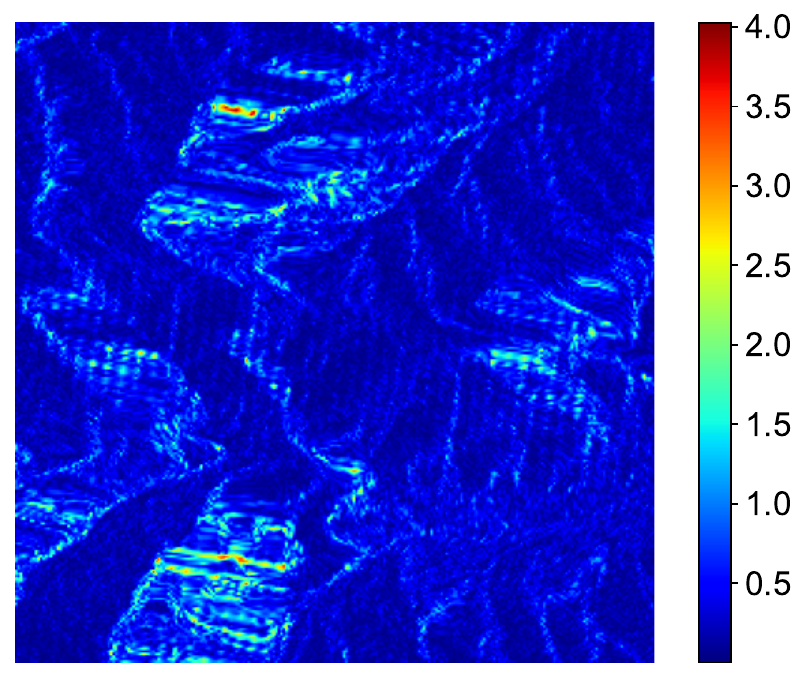}}\hspace{2pt}
    \subfloat[MF]{\includegraphics[width=1.7in]{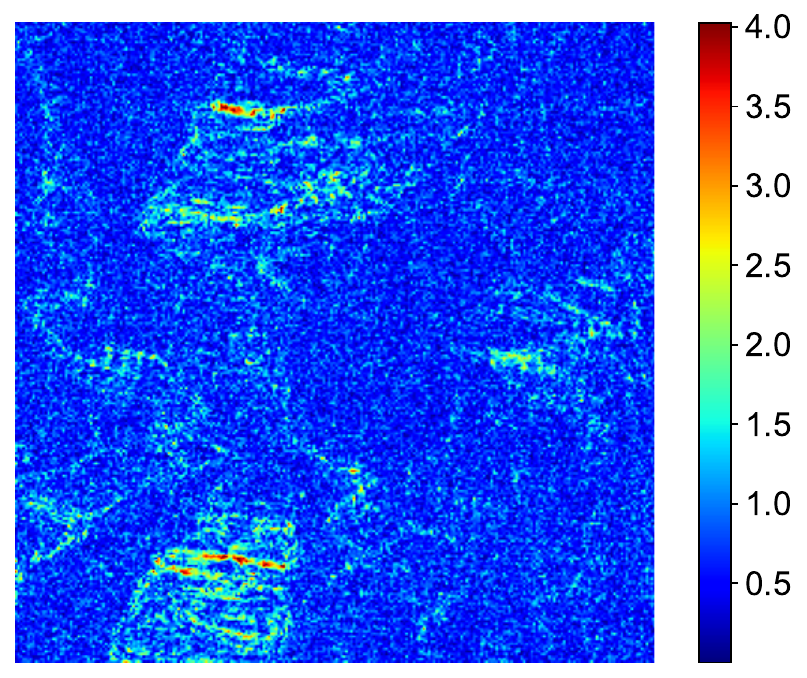}}\hspace{2pt}
    \subfloat[FISTA]{\includegraphics[width=1.7in]{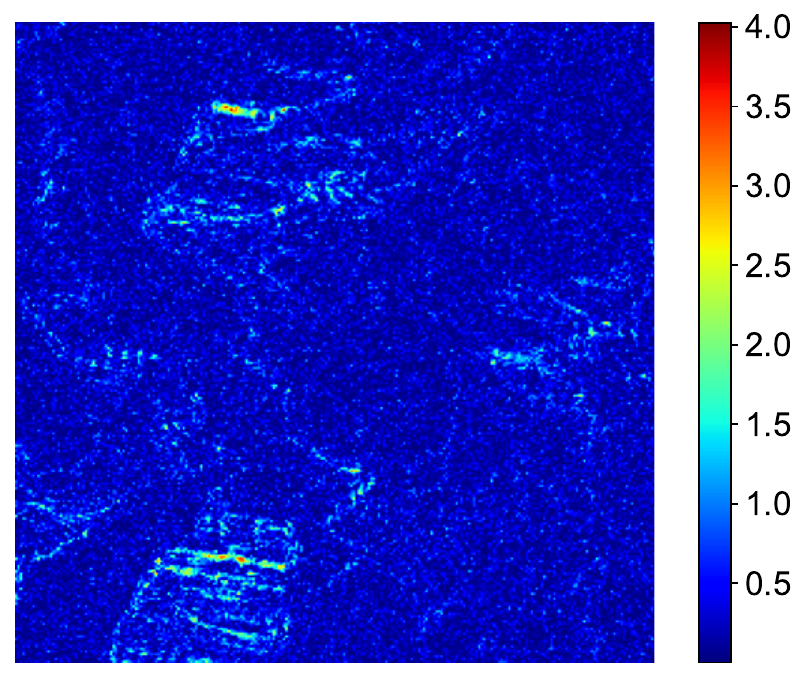}}\hspace{2pt}
    \subfloat[ADMM]{\includegraphics[width=1.7in]{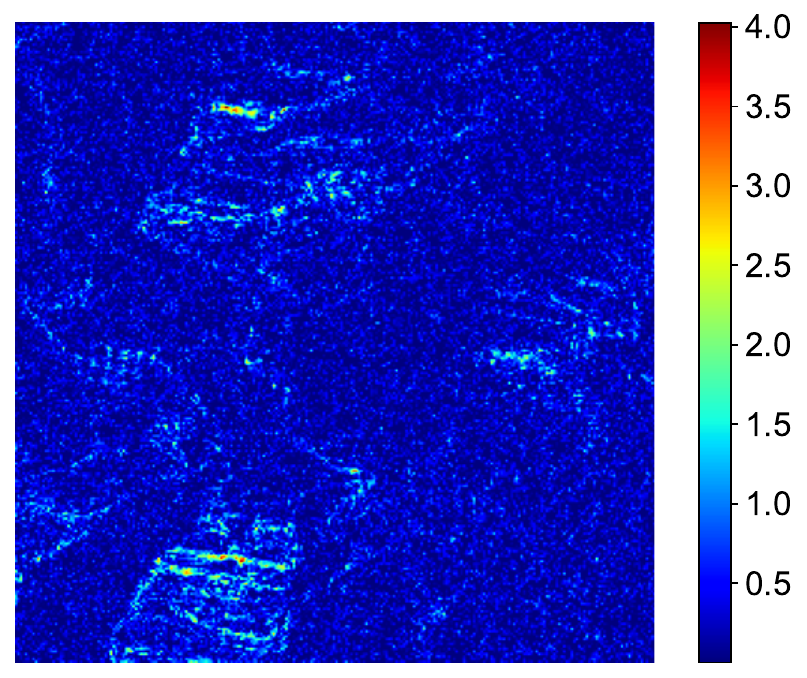}}\hspace{2pt}

    \subfloat[DPS]{\includegraphics[width=1.7in]{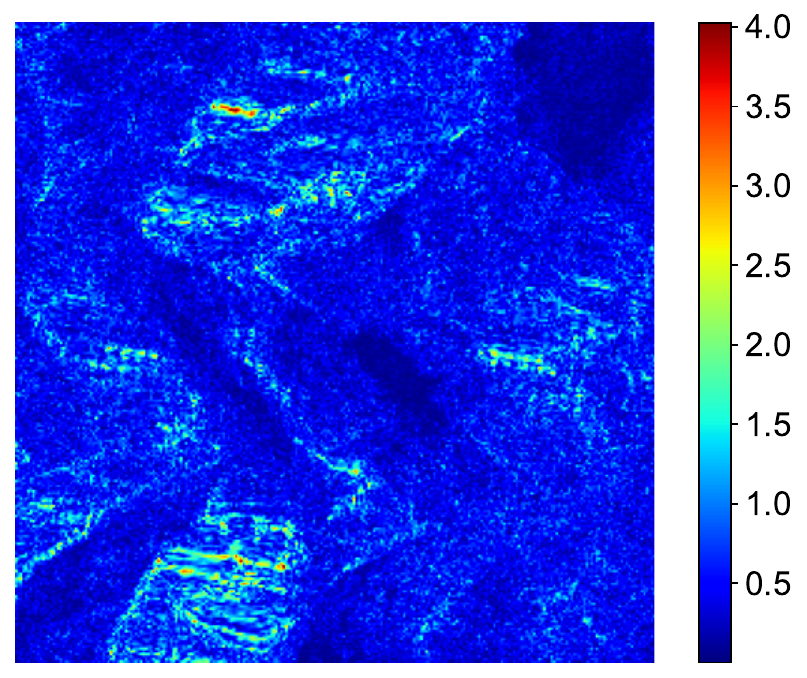}}\hspace{2pt}
    \subfloat[SGS-DDPM]{\includegraphics[width=1.7in]{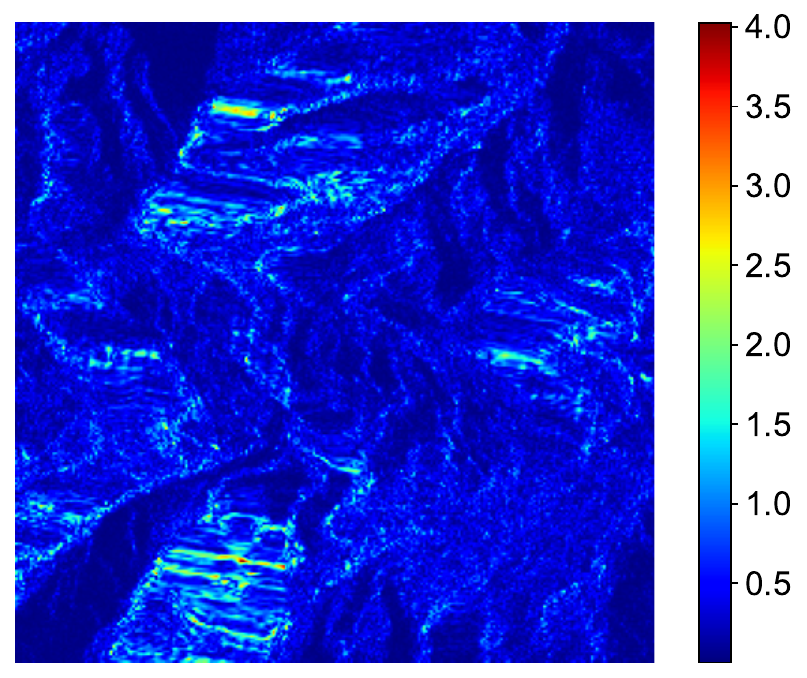}}\hspace{2pt}
    \subfloat[SGS-DDIM]{\includegraphics[width=1.7in]{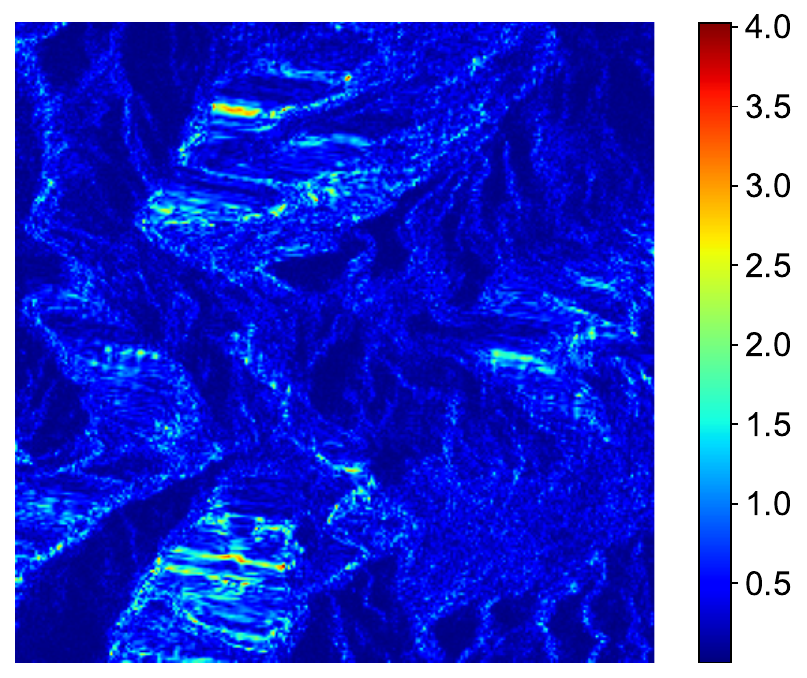}}
    \caption{Reconstruction results comparison under 2 dB SNR and 24 sampling points. Single recovery results of each algorithm.}
    \label{fig:real_scene2_recon_results}
\end{figure*}

Finally, we conducted experiments in two real-world scenes. Scene~I contains relatively few strong-scattering regions, while Scene~II has more such areas. Both scenes include numerous weak-scattering regions rich in fine texture details. Reconstruction was performed under an SNR of 2~dB with 192 sampling points. Fig.~\ref{fig:real_scene1_recon_results} and Fig.~\ref{fig:real_scene2_recon_results} qualitatively present the single-sample reconstruction results of six different methods.

For strong-scattering regions, all six methods achieve a certain degree of reconstruction; however, SGS-based methods produce clearer and more distinct boundaries. Particularly, FISTA and ADMM, due to the sparse prior, recover only a portion of the stronger scattering regions. In the weak-scattering regions containing complex textures, only SGS-based methods successfully preserve detailed structures, while all other methods tend to oversmooth these areas. MF and DPS suffer from high sidelobes, leading to severe texture loss, and sparse reconstruction methods (FISTA and ADMM) also fail to recover these regions effectively.

\begin{table}[!t]
\centering
\caption{Quantitative evaluation of real scene I under 2 dB SNR and 24 sampling points. NMSE and PSNR are reported in dB. \textbf{Bold}:best; \underline{Underline}: second best.}
\begin{tabular}{ccccccc}
\toprule
Method & NMSE $\downarrow$ & PSNR $\uparrow$ & SSIM $\uparrow$ & LPIPS $\downarrow$\\
\midrule
MF      & 2.80     & 19.1    & 0.258    & 0.440\\
FISTA   & \textbf{-1.90}    & 23.7    & 0.404    & 0.421\\
ADMM    & -1.81    & 23.2    & 0.378    & 0.425\\
DPS     & 0.203    & 22.8    & 0.371    & 0.399\\
SGS-DDPM& -1.72   & \underline{25.4}   & \underline{0.479}   & \underline{0.352}\\
SGS-DDIM& \underline{-1.83}   & \textbf{25.5}   & \textbf{0.493}   & \textbf{0.344}\\
\bottomrule
\end{tabular}
\label{tab:real_scene1_quantitative_evaluation}
\end{table}

\begin{table}[!t]
\centering
\caption{Quantitative evaluation of real scene II under 2 dB SNR and 24 sampling points. NMSE and PSNR are reported in dB. \textbf{Bold}:best; \underline{Underline}: second best.}
\begin{tabular}{ccccccc}
\toprule
Method & NMSE $\downarrow$ & PSNR $\uparrow$ & SSIM $\uparrow$ & LPIPS $\downarrow$\\
\midrule
MF      & 2.81     & 17.3    & 0.260    & 0.450\\
FISTA   & \textbf{-1.84}    & 21.9    & 0.370    & 0.430\\
ADMM    & -1.74    & 21.3    & 0.350    & 0.440\\
DPS     & 0.27    & 21.0    & 0.35    & 0.410\\
SGS-DDPM& -1.68   & \underline{23.5}   & \underline{0.430}   & \underline{0.350}\\
SGS-DDIM& \underline{-1.78}   & \textbf{23.6}   & \textbf{0.440}   & \textbf{0.340}\\
\bottomrule
\end{tabular}
\label{tab:real_scene2_quantitative_evaluation}
\end{table}
From the quantitative results in Tables~\ref{tab:real_scene1_quantitative_evaluation} and \ref{tab:real_scene2_quantitative_evaluation}, SGS-DDIM and SGS-DDPM achieve the best or second-best performance in PSNR, SSIM, and LPIPS across both scenes. For NMSE, SGS-DDIM also attains the second-best result, while SGS-DDPM, although slightly behind the sparse reconstruction methods, shows a difference of less than 0.2~dB, indicating excellent reconstruction accuracy and energy preservation. MF performs the worst, highlighting the limitations of reconstruction without prior information under low-sampling and noisy conditions. SGS-DDIM consistently achieves the highest SSIM values, demonstrating its ability to preserve structural and textural similarity to the ground truth, whereas DPS is less effective in weak-scattering regions despite guidance from the diffusion prior. In terms of LPIPS, SGS-based methods outperform all others, reflecting superior perceptual quality and detail preservation.

\section{Conclusion}\label{conclusion}
In this paper, we extended a diffusion-driven split Gibbs sampling framework to SAR imaging, which integrates a pretrained diffusion model within a proximal sampling scheme. By alternately performing likelihood-constrained updates and prior-driven sampling, the method ensures progressive convergence toward the posterior distribution, effectively leveraging high-order learned priors while rigorously enforcing measurement fidelity. Compared with existing diffusion-based SAR reconstruction approaches, this framework overcomes the limitations of oversimplified likelihood approximations and weakened prior regularization. Experiments on simulated and Sentinel-1A datasets demonstrate over 7~dB average PSNR improvement in simulations, significant sidelobe suppression (MPLSR +2.96~dB, MISLR +11.5~dB), and over 1.6~dB PSNR gain on real-world data. The method accurately recovers fine structures, including ridges, edges, and textures, while effectively suppressing noise and artifacts, producing sharper, more realistic reconstructions.

\bibliographystyle{IEEEtran}
\bibliography{reference}

\begin{thebibliography}{10}
\providecommand{\url}[1]{#1}
\csname url@samestyle\endcsname
\providecommand{\newblock}{\relax}
\providecommand{\bibinfo}[2]{#2}
\providecommand{\BIBentrySTDinterwordspacing}{\spaceskip=0pt\relax}
\providecommand{\BIBentryALTinterwordstretchfactor}{4}
\providecommand{\BIBentryALTinterwordspacing}{\spaceskip=\fontdimen2\font plus
\BIBentryALTinterwordstretchfactor\fontdimen3\font minus \fontdimen4\font\relax}
\providecommand{\BIBforeignlanguage}[2]{{%
\expandafter\ifx\csname l@#1\endcsname\relax
\typeout{** WARNING: IEEEtran.bst: No hyphenation pattern has been}%
\typeout{** loaded for the language `#1'. Using the pattern for}%
\typeout{** the default language instead.}%
\else
\language=\csname l@#1\endcsname
\fi
#2}}
\providecommand{\BIBdecl}{\relax}
\BIBdecl

\bibitem{kovaly1976synthetic}
J.~Kovaly, \emph{Synthetic aperture radar}.\hskip 1em plus 0.5em minus 0.4em\relax Artech House, Incorporated, 1976.

\bibitem{cumming2005digital}
I.~Cumming and F.~Wong, \emph{Digital Processing of Synthetic Aperture Radar Data: Algorithms and Implementation}.\hskip 1em plus 0.5em minus 0.4em\relax Artech House, 2005, no.~1.

\bibitem{osti_293027}
F.~M. Henderson and A.~J. Lewis, \emph{Principles and applications of imaging radar. Manual of remote sensing: Third edition, Volume 2}.\hskip 1em plus 0.5em minus 0.4em\relax John Wiley and Sons, Inc., Somerset, NJ (United States), 12 1998.

\bibitem{zhang2024iterative}
Y.~Zhang, X.~Bu, X.~Ge, J.~Xin, and X.~Liang, ``An iterative motion compensation algorithm for synthetic aperture passive positioning,'' \emph{IEEE Signal Processing Letters}, 2024.

\bibitem{chen2025data}
C.~Chen, H.~Yang, S.~Chen, F.~Xi, and Z.~Liu, ``A data-driven motion compensation scheme for compressed sensing sar image restoration,'' \emph{IEEE Transactions on Geoscience and Remote Sensing}, 2025.

\bibitem{kang2023sar}
M.-S. Kang and J.-M. Baek, ``Sar image reconstruction via incremental imaging with compressive sensing,'' \emph{IEEE Transactions on Aerospace and Electronic Systems}, vol.~59, no.~4, pp. 4450--4463, 2023.

\bibitem{wang2023atasi}
M.~Wang, Z.~Zhang, X.~Qiu, S.~Gao, and Y.~Wang, ``Atasi-net: An efficient sparse reconstruction network for tomographic sar imaging with adaptive threshold,'' \emph{IEEE Transactions on Geoscience and Remote Sensing}, vol.~61, pp. 1--18, 2023.

\bibitem{farhadiani2022sar}
R.~Farhadiani, S.~Homayouni, A.~Bhattacharya, and M.~Mahdianpari, ``Sar despeckling based on cnn and bayesian estimator in complex wavelet domain,'' \emph{IEEE Geoscience and Remote Sensing Letters}, vol.~19, pp. 1--5, 2022.

\bibitem{wang2025diffusion}
Z.~Wang, J.~Han, and C.~Zhang, ``Diffusion posterior sampling for sar despeckling,'' \emph{IEEE Transactions on Geoscience and Remote Sensing}, 2025.

\bibitem{zhang2023pnp}
H.~Zhang, J.~Ni, C.~Zhang, Y.~Luo, and Q.~Zhang, ``Pnp-based ground moving target imaging network for squint sar and sparse sampling,'' \emph{IEEE Transactions on Geoscience and Remote Sensing}, vol.~62, pp. 1--20, 2023.

\bibitem{jiang2025high}
N.~Jiang, J.~Chen, J.~Zhu, B.~Liang, D.~Yang, X.~Huang, and M.~Xing, ``High frame rate along-track swarm sar sub-aperture collaboration imaging for moving target,'' \emph{IEEE Transactions on Geoscience and Remote Sensing}, 2025.

\bibitem{hui2025bidirectional}
X.~Hui, Z.~Liu, L.~Wang, Z.~Zhang, and S.~Yao, ``Bidirectional interaction fusion network based on ec-maps and sar images for sar target recognition,'' \emph{IEEE Transactions on Instrumentation and Measurement}, 2025.

\bibitem{ni2024dpgunet}
K.~Ni, C.~Yuan, Z.~Zheng, N.~Huang, and P.~Wang, ``Dpgunet: A dynamic pyramidal graph u-net for sar image classification,'' \emph{IEEE Transactions on Aerospace and Electronic Systems}, vol.~60, no.~4, pp. 5247--5263, 2024.

\bibitem{pena2024deepaqua}
F.~J. Pe{\~n}a, C.~H{\"u}binger, A.~H. Payberah, and F.~Jaramillo, ``Deepaqua: Semantic segmentation of wetland water surfaces with sar imagery using deep neural networks without manually annotated data,'' \emph{International Journal of Applied Earth Observation and Geoinformation}, vol. 126, p. 103624, 2024.

\bibitem{deng2024hyperspectral}
B.~Deng, P.~Duan, X.~Lu, Z.~Wang, and X.~Kang, ``Hyperspectral and sar image classification via graph convolutional fusion network,'' \emph{IEEE Transactions on Geoscience and Remote Sensing}, 2024.

\bibitem{gu2025hpn}
P.~Gu, W.~Liu, S.~Feng, T.~Wei, J.~Wang, and H.~Chen, ``Hpn-cr: Heterogeneous parallel network for sar-optical data fusion cloud removal,'' \emph{IEEE Transactions on Geoscience and Remote Sensing}, 2025.

\bibitem{bamler2002comparison}
R.~Bamler, ``A comparison of range-doppler and wavenumber domain sar focusing algorithms,'' \emph{IEEE Transactions on Geoscience and Remote Sensing}, vol.~30, no.~4, pp. 706--713, 2002.

\bibitem{moreira2002extended}
A.~Moreira, J.~Mittermayer, and R.~Scheiber, ``Extended chirp scaling algorithm for air-and spaceborne sar data processing in stripmap and scansar imaging modes,'' \emph{IEEE Transactions on geoscience and remote sensing}, vol.~34, no.~5, pp. 1123--1136, 2002.

\bibitem{cumming2003interpretations}
I.~G. Cumming, Y.~L. Neo, and F.~H. Wong, ``Interpretations of the omega-k algorithm and comparisons with other algorithms,'' in \emph{IGARSS 2003. 2003 IEEE International Geoscience and Remote Sensing Symposium. Proceedings (IEEE Cat. No. 03CH37477)}, vol.~3.\hskip 1em plus 0.5em minus 0.4em\relax IEEE, 2003, pp. 1455--1458.

\bibitem{yegulalp1999fast}
A.~F. Yegulalp, ``Fast backprojection algorithm for synthetic aperture radar,'' in \emph{Proceedings of the 1999 IEEE Radar Conference. Radar into the Next Millennium (Cat. No. 99CH36249)}.\hskip 1em plus 0.5em minus 0.4em\relax IEEE, 1999, pp. 60--65.

\bibitem{candes2006robust}
E.~J. Cand{\`e}s, J.~Romberg, and T.~Tao, ``Robust uncertainty principles: Exact signal reconstruction from highly incomplete frequency information,'' \emph{IEEE Transactions on information theory}, vol.~52, no.~2, pp. 489--509, 2006.

\bibitem{candes2006near}
E.~J. Candes and T.~Tao, ``Near-optimal signal recovery from random projections: Universal encoding strategies?'' \emph{IEEE transactions on information theory}, vol.~52, no.~12, pp. 5406--5425, 2006.

\bibitem{eldar2012compressed}
Y.~C. Eldar and G.~Kutyniok, \emph{Compressed sensing: theory and applications}.\hskip 1em plus 0.5em minus 0.4em\relax Cambridge university press, 2012.

\bibitem{eldar2015sampling}
Y.~C. Eldar, \emph{Sampling theory: Beyond bandlimited systems}.\hskip 1em plus 0.5em minus 0.4em\relax Cambridge University Press, 2015.

\bibitem{yang2013segmented}
J.~Yang, J.~Thompson, X.~Huang, T.~Jin, and Z.~Zhou, ``Segmented reconstruction for compressed sensing sar imaging,'' \emph{IEEE transactions on geoscience and remote sensing}, vol.~51, no.~7, pp. 4214--4225, 2013.

\bibitem{aberman2017sub}
K.~Aberman and Y.~C. Eldar, ``Sub-nyquist sar via fourier domain range-doppler processing,'' \emph{IEEE Transactions on Geoscience and Remote Sensing}, vol.~55, no.~11, pp. 6228--6244, 2017.

\bibitem{de2019compressed}
A.~De~Maio, Y.~C. Eldar, and A.~M. Haimovich, \emph{Compressed sensing in radar signal processing}.\hskip 1em plus 0.5em minus 0.4em\relax Cambridge University Press, 2019.

\bibitem{hu2021compressive}
X.~Hu, C.~Ma, X.~Lu, and T.~S. Yeo, ``Compressive sensing sar imaging algorithm for lfmcw systems,'' \emph{IEEE Transactions on Geoscience and Remote Sensing}, vol.~59, no.~10, pp. 8486--8500, 2021.

\bibitem{kang2025compressive}
M.-S. Kang and J.-M. Baek, ``Compressive sensing based omega-k algorithm for sar focusing,'' \emph{IEEE Geoscience and Remote Sensing Letters}, 2025.

\bibitem{beck2009fast}
A.~Beck and M.~Teboulle, ``A fast iterative shrinkage-thresholding algorithm for linear inverse problems,'' \emph{SIAM journal on imaging sciences}, vol.~2, no.~1, pp. 183--202, 2009.

\bibitem{palomar2010convex}
D.~P. Palomar and Y.~C. Eldar, \emph{Convex optimization in signal processing and communications}.\hskip 1em plus 0.5em minus 0.4em\relax Cambridge university press, 2010.

\bibitem{boyd2011distributed}
S.~Boyd, N.~Parikh, E.~Chu, B.~Peleato, J.~Eckstein \emph{et~al.}, ``Distributed optimization and statistical learning via the alternating direction method of multipliers,'' \emph{Foundations and Trends{\textregistered} in Machine learning}, vol.~3, no.~1, pp. 1--122, 2011.

\bibitem{needell2009uniform}
D.~Needell and R.~Vershynin, ``Uniform uncertainty principle and signal recovery via regularized orthogonal matching pursuit,'' \emph{Foundations of computational mathematics}, vol.~9, no.~3, pp. 317--334, 2009.

\bibitem{tipping2001sparse}
M.~E. Tipping, ``Sparse bayesian learning and the relevance vector machine,'' \emph{Journal of machine learning research}, vol.~1, no. Jun, pp. 211--244, 2001.

\bibitem{wipf2004sparse}
D.~P. Wipf and B.~D. Rao, ``Sparse bayesian learning for basis selection,'' \emph{IEEE Transactions on Signal processing}, vol.~52, no.~8, pp. 2153--2164, 2004.

\bibitem{xu2021image}
W.~Xu, B.~Wang, M.~Xiang, R.~Li, and W.~Li, ``Image defocus in an airborne uwb vhr microwave photonic sar: Analysis and compensation,'' \emph{IEEE Transactions on Geoscience and Remote Sensing}, vol.~60, pp. 1--18, 2021.

\bibitem{dong2022high}
B.~Dong, G.~Li, and Q.~Zhang, ``High-resolution and wide-swath imaging of spaceborne sar via random prf variation constrained by the coverage diagram,'' \emph{IEEE Transactions on Geoscience and Remote Sensing}, vol.~60, pp. 1--16, 2022.

\bibitem{bonfert2024improving}
C.~Bonfert, E.~Ruopp, and C.~Waldschmidt, ``Improving sar imaging by superpixel-based compressed sensing and backprojection processing,'' \emph{IEEE Transactions on Geoscience and Remote Sensing}, vol.~62, pp. 1--12, 2024.

\bibitem{an2021geosynchronous}
H.~An, J.~Wu, K.~C. Teh, Z.~Sun, and J.~Yang, ``Geosynchronous spaceborne--airborne bistatic sar imaging based on fast low-rank and sparse matrices recovery,'' \emph{IEEE Transactions on Geoscience and Remote Sensing}, vol.~60, pp. 1--14, 2021.

\bibitem{an2021joint}
H.~An, J.~Wu, K.~C. Teh, Z.~Sun, Z.~Li, and J.~Yang, ``Joint low-rank and sparse tensors recovery for video synthetic aperture radar imaging,'' \emph{IEEE Transactions on Geoscience and Remote Sensing}, vol.~60, pp. 1--13, 2021.

\bibitem{bi2021sparse}
H.~Bi, X.~Lu, Y.~Yin, W.~Yang, and D.~Zhu, ``Sparse sar imaging based on periodic block sampling data,'' \emph{IEEE Transactions on Geoscience and Remote Sensing}, vol.~60, pp. 1--12, 2021.

\bibitem{fan2024integrating}
Y.~Fan, B.~Zhang, and Y.~Wu, ``Integrating regularization and pnp priors for sar image reconstruction using multi-agent consensus equilibrium,'' \emph{IEEE Transactions on Geoscience and Remote Sensing}, 2024.

\bibitem{yang2024sar}
J.~Yang, H.~Zuo, H.~An, R.~Jiang, Z.~Li, Z.~Sun, and J.~Wu, ``Sar nonsparse scene reconstruction network via image feature representation learning,'' \emph{IEEE Transactions on Geoscience and Remote Sensing}, vol.~62, pp. 1--15, 2024.

\bibitem{li2022stls}
M.~Li, J.~Wu, W.~Huo, Z.~Li, J.~Yang, and H.~Li, ``Stls-ladmm-net: A deep network for sar autofocus imaging,'' \emph{IEEE Transactions on Geoscience and Remote Sensing}, vol.~60, pp. 1--14, 2022.

\bibitem{xiong2021q}
K.~Xiong, G.~Zhao, Y.~Wang, G.~Shi, and S.~Chen, ``L q-spb-net: A real-time deep network for sar imaging and despeckling,'' \emph{IEEE Transactions on Geoscience and Remote Sensing}, vol.~60, pp. 1--21, 2021.

\bibitem{xu2024joint}
Y.~Xu, X.~Zhang, S.~Wei, J.~Shi, T.~Zeng, X.~Xu, W.~Zhang, and X.~Zhan, ``Joint generalized lq and convolutional regularization: Enhancing mmw automotive sar sparse imaging,'' \emph{IEEE Transactions on Geoscience and Remote Sensing}, 2024.

\bibitem{wu2024mf}
Y.~Wu, Z.~Zhang, X.~Qiu, Y.~Zhao, and W.~Yu, ``Mf-jmodl-net: A sparse sar imaging network for undersampling pattern design toward suppressed azimuth ambiguity,'' \emph{IEEE Transactions on Geoscience and Remote Sensing}, vol.~62, pp. 1--18, 2024.

\bibitem{li2024sar}
M.~Li, W.~Huo, J.~Wu, and J.~Yang, ``Sar image reconstruction method for target detection using self-attention cnn-based deep prior learning,'' \emph{IEEE Transactions on Geoscience and Remote Sensing}, 2024.

\bibitem{wang2024synthetic}
Z.~Wang, C.~Song, Z.~Jiao, B.~Wang, and M.~Xiang, ``Synthetic aperture radar deep statistical imaging through diffusion-generaive-model conditional inference,'' \emph{IEEE Transactions on Geoscience and Remote Sensing}, 2024.

\bibitem{chung2022diffusion}
H.~Chung, J.~Kim, M.~T. Mccann, M.~L. Klasky, and J.~C. Ye, ``Diffusion posterior sampling for general noisy inverse problems,'' \emph{arXiv preprint arXiv:2209.14687}, 2022.

\bibitem{xu2024provably}
X.~Xu and Y.~Chi, ``Provably robust score-based diffusion posterior sampling for plug-and-play image reconstruction,'' \emph{Advances in Neural Information Processing Systems}, vol.~37, pp. 36\,148--36\,184, 2024.

\bibitem{8344566}
H.~Bi, G.~Bi, B.~Zhang, and W.~Hong, ``Complex-image-based sparse sar imaging and its equivalence,'' \emph{IEEE Transactions on Geoscience and Remote Sensing}, vol.~56, no.~9, pp. 5006--5014, 2018.

\bibitem{dong2014sar}
X.~Dong and Y.~Zhang, ``Sar image reconstruction from undersampled raw data using maximum a posteriori estimation,'' \emph{IEEE Journal of Selected Topics in Applied Earth Observations and Remote Sensing}, vol.~8, no.~4, pp. 1651--1664, 2014.

\bibitem{kelly2012advanced}
S.~I. Kelly, C.~Du, G.~Rilling, and M.~E. Davies, ``Advanced image formation and processing of partial synthetic aperture radar data,'' \emph{IET signal processing}, vol.~6, no.~5, pp. 511--520, 2012.

\bibitem{patel2010compressed}
V.~M. Patel, G.~R. Easley, D.~M. Healy, and R.~Chellappa, ``Compressed synthetic aperture radar,'' \emph{IEEE Journal of selected topics in signal processing}, vol.~4, no.~2, pp. 244--254, 2010.

\bibitem{welling2011bayesian}
M.~Welling and Y.~W. Teh, ``Bayesian learning via stochastic gradient langevin dynamics,'' in \emph{Proceedings of the 28th international conference on machine learning (ICML-11)}, 2011, pp. 681--688.

\bibitem{hastings1970monte}
W.~K. Hastings, ``Monte carlo sampling methods using markov chains and their applications,'' 1970.

\bibitem{song2019generative}
Y.~Song and S.~Ermon, ``Generative modeling by estimating gradients of the data distribution,'' \emph{Advances in neural information processing systems}, vol.~32, 2019.

\bibitem{song2020score}
Y.~Song, J.~Sohl-Dickstein, D.~P. Kingma, A.~Kumar, S.~Ermon, and B.~Poole, ``Score-based generative modeling through stochastic differential equations,'' \emph{arXiv preprint arXiv:2011.13456}, 2020.

\bibitem{sohl2015deep}
J.~Sohl-Dickstein, E.~Weiss, N.~Maheswaranathan, and S.~Ganguli, ``Deep unsupervised learning using nonequilibrium thermodynamics,'' in \emph{International conference on machine learning}.\hskip 1em plus 0.5em minus 0.4em\relax pmlr, 2015, pp. 2256--2265.

\bibitem{luo2022understanding}
C.~Luo, ``Understanding diffusion models: A unified perspective,'' \emph{arXiv preprint arXiv:2208.11970}, 2022.

\bibitem{vono2019split}
M.~Vono, N.~Dobigeon, and P.~Chainais, ``Split-and-augmented gibbs sampler—application to large-scale inference problems,'' \emph{IEEE Transactions on Signal Processing}, vol.~67, no.~6, pp. 1648--1661, 2019.

\bibitem{10541919}
F.~Coeurdoux, N.~Dobigeon, and P.~Chainais, ``Plug-and-play split gibbs sampler: Embedding deep generative priors in bayesian inference,'' \emph{IEEE Transactions on Image Processing}, vol.~33, pp. 3496--3507, 2024.

\bibitem{zhang2022fast}
Q.~Zhang and Y.~Chen, ``Fast sampling of diffusion models with exponential integrator,'' \emph{arXiv preprint arXiv:2204.13902}, 2022.

\bibitem{song2020denoising}
J.~Song, C.~Meng, and S.~Ermon, ``Denoising diffusion implicit models,'' \emph{arXiv preprint arXiv:2010.02502}, 2020.

\bibitem{tian2025dic}
Y.~Tian, J.~Han, C.~Wang, Y.~Liang, C.~Xu, and H.~Chen, ``Dic: Rethinking conv3x3 designs in diffusion models,'' in \emph{Proceedings of the Computer Vision and Pattern Recognition Conference}, 2025, pp. 2469--2478.

\bibitem{zhang2018unreasonable}
R.~Zhang, P.~Isola, A.~A. Efros, E.~Shechtman, and O.~Wang, ``The unreasonable effectiveness of deep features as a perceptual metric,'' in \emph{Proceedings of the IEEE conference on computer vision and pattern recognition}, 2018, pp. 586--595.

\bibitem{lecun2002gradient}
Y.~LeCun, L.~Bottou, Y.~Bengio, and P.~Haffner, ``Gradient-based learning applied to document recognition,'' \emph{Proceedings of the IEEE}, vol.~86, no.~11, pp. 2278--2324, 2002.

\end{thebibliography}


 





\end{document}